\documentclass[aps,pre,twocolumn]{revtex4-1}
\usepackage{graphicx}
\usepackage{dcolumn}
\usepackage{bm}
\usepackage{bbm}

\usepackage{amsmath}
\usepackage{amssymb}
\usepackage{verbatim}

\begin{document}

\title{Covariance matrix filtering with bootstrapped hierarchies}

\author{Christian Bongiorno}
\affiliation{Universit\'e Paris-Saclay, CentraleSup\'elec, Laboratoire de Math\'ematiques et Informatique pour les Syst\`emes Complexes, 91190, Gif-sur-Yvette, France}
\author{Damien Challet}
\affiliation{Universit\'e Paris-Saclay, CentraleSup\'elec, Laboratoire de Math\'ematiques et Informatique pour les Syst\`emes Complexes, 91190, Gif-sur-Yvette, France}

\begin{abstract}
Cleaning covariance matrices is a highly non-trivial problem, yet of central importance in the statistical inference of dependence between objects. We propose here a probabilistic hierarchical clustering method, named Bootstrapped Average Hierarchical Clustering (BAHC) that is particularly effective in the high-dimensional case, i.e., when there are more objects than features. When applied to DNA microarray, our method yields distinct hierarchical structures that cannot be accounted for by usual hierarchical clustering. We then use global minimum-variance risk management  to test our method and find that BAHC leads to significantly smaller realized risk compared to state-of-the-art linear and nonlinear filtering methods in the high-dimensional case. Spectral decomposition shows that BAHC better captures the persistence of the dependence structure between asset price returns in the calibration and the test periods.
\end{abstract}

\maketitle

Covariance matrix inference is a cornerstone of the dependence inference between objects. This kind of matrix suffers however from the curse of dimensionality, as they become very noisy when the number of objects is similar to the number of features. Even worse, unfiltered covariance matrices are pathological in the high dimensional case, i.e., when the number of features exceeds the number of objects. This case is frequent e.g. in biological data and in multivariate dynamical systems such as financial markets in which only the most recent history is likely to be relevant.

Given its importance, covariance matrix filtering has a long history.  A popular approach is to obtain filtered covariance matrices from the corresponding correlation matrices.
Two types of approaches stand out: $i)$ spectral methods, e.g. Random Matrix Theory, Rotationally Invariant Estimators ~\cite{bun2017cleaning}, and Shrinkage~\cite{ledoit2004well,ledoit2017nonlinear}; $ii)$ ansatz for the correlation matrix, e.g. block-diagonal~\cite{beguvsic2019cluster} or hierarchical~\cite{tumminello2007hierarchically}.

The usual setting is to have $n$ objects and $t$ features and to compute the correlation matrix between these $n$ objects. Recent results on Rotationally Invariant Estimators~\cite{bun2016rotational} propose non-linear shrinkage methods able to correct the eigenvalue spectrum of covariance matrices optimally: the inversion of the QuEST function~\cite{ledoit2012nonlinear}, the Cross-Validated (CV) eigenvalue shrinkage~\cite{bartz2016cross} and the IW-regularization~\cite{bun2017cleaning}, the latter being valid only in the low dimensional regime $q=n/t<1$, i.e., when there are more features than objects. 
Eigenvector filtering is more complex. However, ans\"atze for the shape of the true correlation matrix impose constraints on the structure of the eigenvectors and of the eigenvalues. Such ansatz should be simple enough to clean noise but flexible enough to account for fine relevant details. The popular hierarchical clustering ansatz (HC thereafter) is indeed simple: it assumes that correlations are nested \cite{tumminello2007hierarchically}, which is equivalent to assume that dependencies are described by a dendrogram (a tree). In practice, it is hard to find statistically-validated hierarchical structures~\cite{bongiorno2019nested} when the fitted hierarchical structure is highly sensitive to small variations of data.  

\begin{figure}

\label{fig:luscgenesamp}\includegraphics[width=0.5\columnwidth]{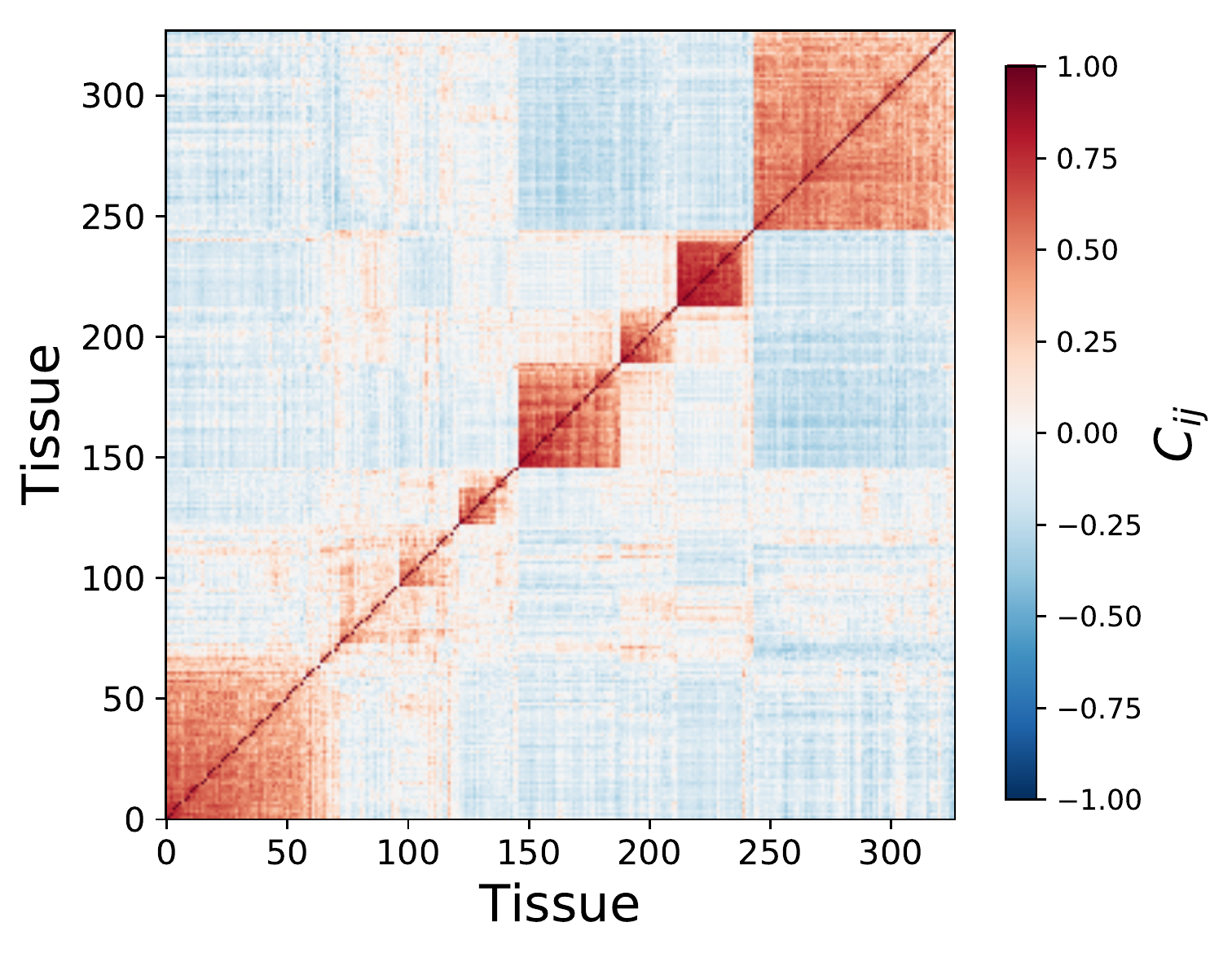}
\label{fig:luscgeneav}\includegraphics[width=0.45\columnwidth]{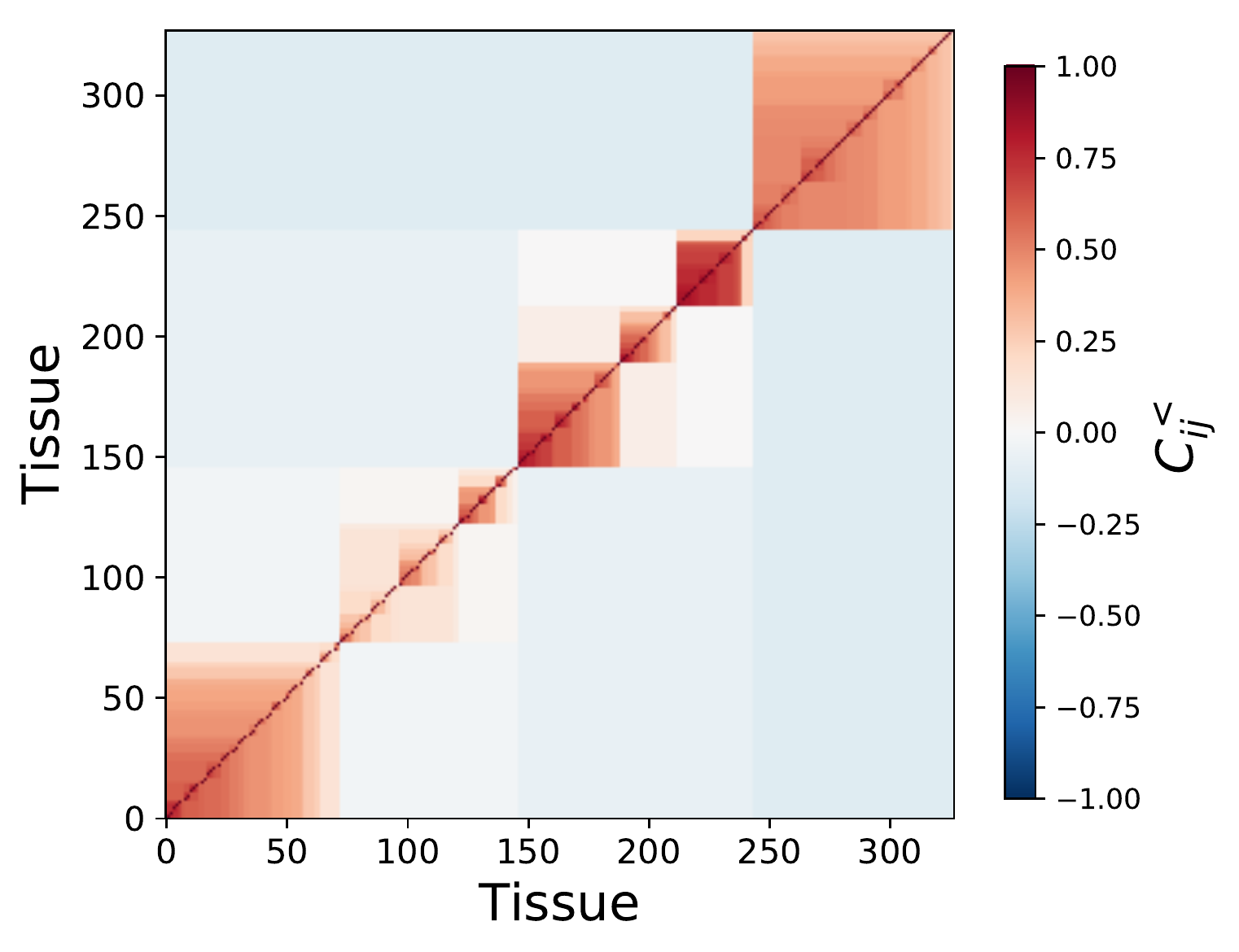}
\label{fig:luscgenebav}\includegraphics[width=0.45\columnwidth]{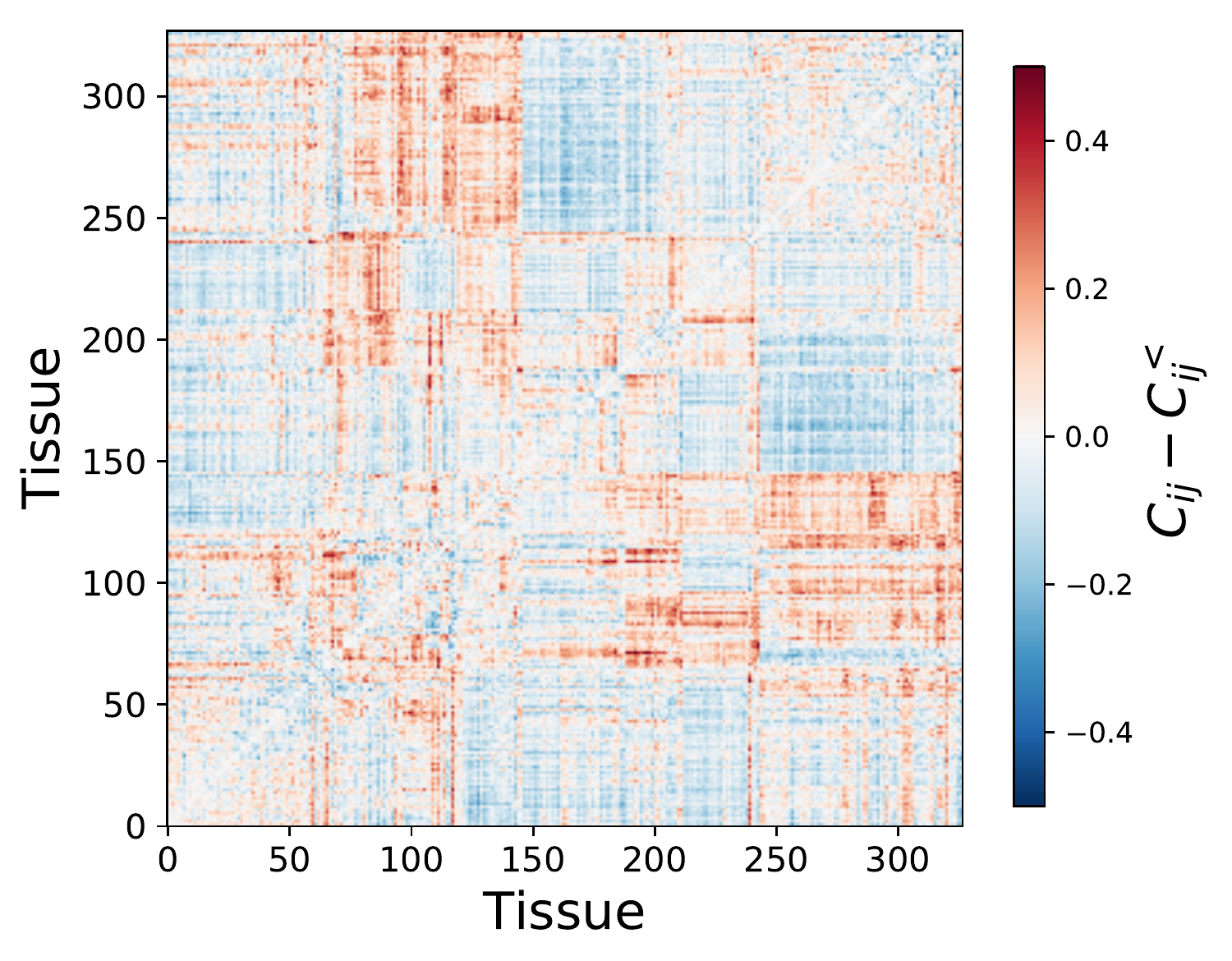}

\caption{Correlation matrix from tissue-gene micro-array data of patients affected by lung cancer. The upper left plot is the sample correlation matrix, the upper right plot is the result of hierarchical and average-linkage averaging (HCAL). The bottom  plot is the difference between the two: it still has evident structure unaccounted for by HCAL.}\label{fig:lung}
\end{figure}

\begin{figure}
\includegraphics[width=0.45\columnwidth]{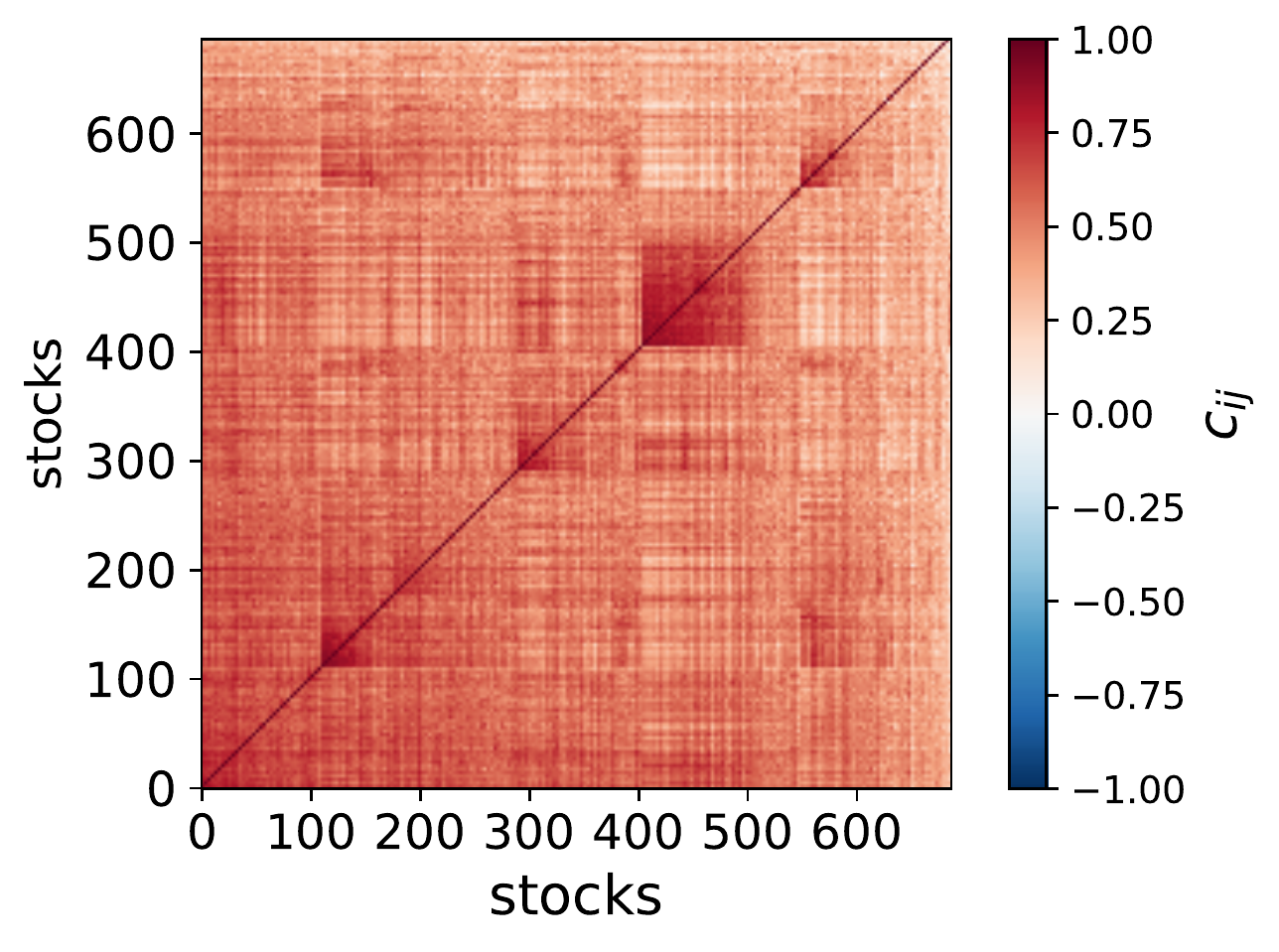} 
\includegraphics[width=0.45\columnwidth]{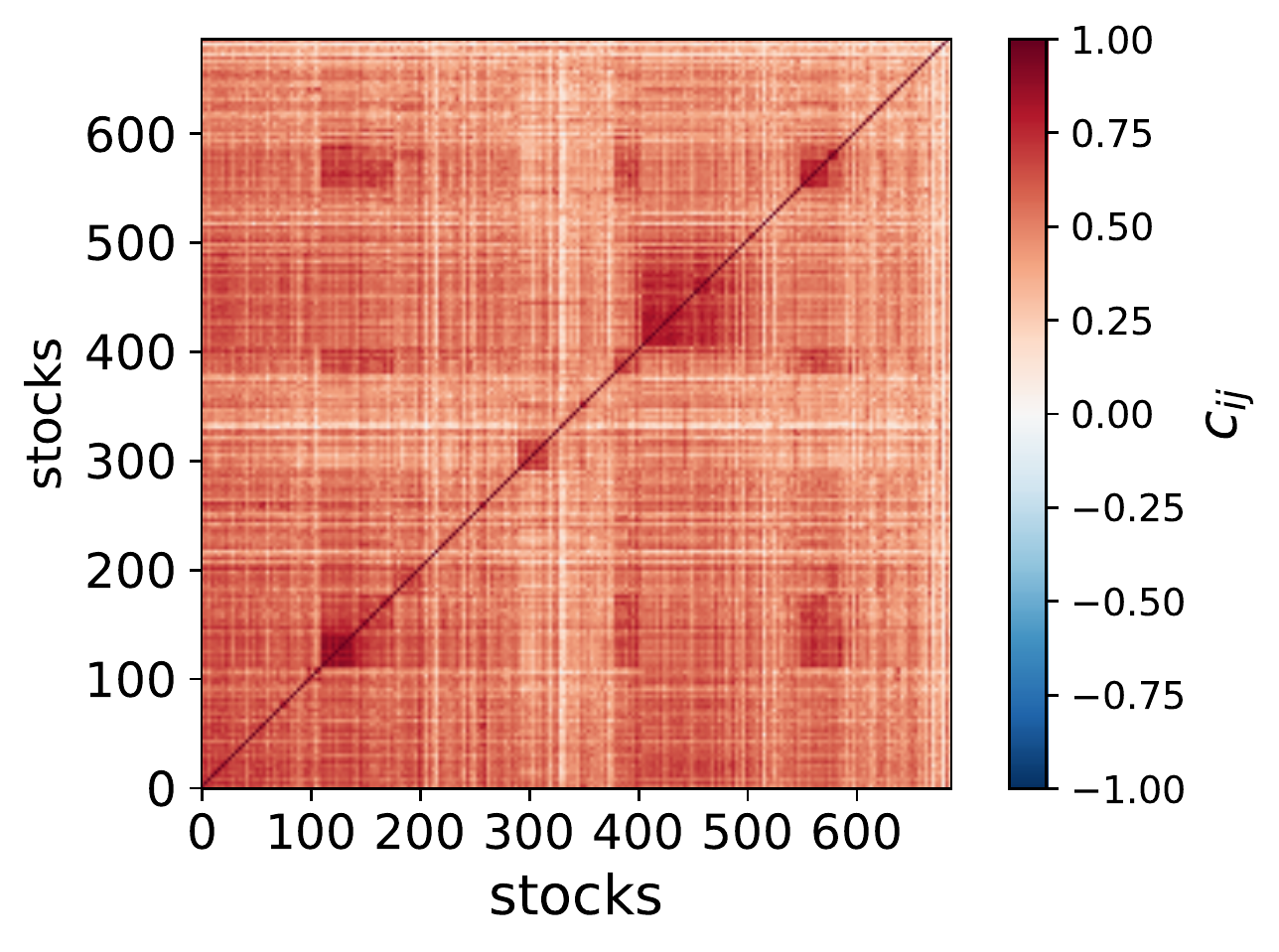} 
\caption{Correlation matrix of US equities price returns in the 2008-01-23 to 2008-11-04 (left plot) and in the 2008-11-05 to 2009-08-24 period (right plot). The elements of both panels are ordered according to the in-sample HCAL dendrogram of the first period.}\label{fig:overlap}
\end{figure}

An obvious problem of HC occurs when the structure is more complex than a tree: for example the non-diagonal blocks in Figs~\ref{fig:lung} and \ref{fig:overlap} are ignored by a hierarchical ansatz: one needs more than a single hierarchical structure to describe these empirical dependence structures. As a consequence, a non-negligible part of the dependence structure is left out, and in a dynamic context, the stability of a single hierarchical structure is likely to be poor.

Here, we introduce a more flexible hierarchical ansatz able to capture more of the structure of the eigenvectors. The rationale is to compute filtered hierarchical structures of many bootstrapped copies of the initial data, which yields probabilistic hierarchical structures. Such procedure describes the structure of correlation and covariance matrices better while keeping the robustness of hierarchical clustering.
 We illustrate the power of our method with data from two relevant fields. First, in bioinformatics, DNA micro-array gene expression dependence in tissues is frequently characterized by correlation matrices. Hierarchical clustering and its variants are commonly used  \cite{quackenbush2001computational,hira2015review}, which helps simplify the covariance matrix by linkage averaging\ \cite{friedman2001elements} (see Fig. \ref{fig:lung}). When there are several strong candidates of hierarchical structure, this approach selects a single one, which neglects possibly crucial information held by alternative structures.  
Comparing unfiltered correlation matrices with the filtering yielded by hierarchical clustering and average linkage (HCAL) \cite{tumminello2007hierarchically} (Fig.\ \ref{fig:lung}) makes it clear first that (i) hierarchical clustering does capture some of the structure and (ii) a substantial part of the structure is lost (see the bottom plot). 
This is because hierarchical clustering imposes too strict a structure, which erases out an uncontrolled amount of information. 

Another domain in which covariance matrix filtering plays a central role is risk management. Broadly speaking, the problem amounts to minimize future uncertainty by determining the fraction of resources to allocate to every possible choice. Risk in this particular context is due to fluctuations of the future value of the choices. The usual procedure consists in minimizing a suitable risk measure in the calibration window and hoping that the future, realized, risk will bear some relationship with the calibrated risk. 

The simplest approach consists in defining risk as the variance of the weighted sum of choices' values and to minimise it. This is known as globaly minimum-variance portfolios, a subfield of quadratic portfolio optimization which has a wide range of applications: investment into technologies\ \cite{hubbard2014measure}, energy sources mix for countries \cite{roques2008fuel,arnesano2012extension}, wind farm locations\ \cite{dunlop2004wind}, and capital allocation in finance \cite{markowitz2000meanvar}. We shall focus on financial risk because data are abundant, which makes it possible to compare the out-of-sample performance of filtering methods. In addition, the high-dimensional regime is particularly relevant in finance: there are many assets to choose from and the speed with which the dependence structure between asset price returns may change asks for an as short as possible calibration period \cite{bongiorno2019nonparametric}. 

In an inference or descriptive context such as DNA microarray data analysis, filtering correlation matrices is meant to bring estimated covariance matrices closer to the ground truth. In a dynamical context, especially for non-stationary systems such as financial markets, what matters is the part of the ground truth that most likely persists after the calibration period, i.e., when one uses the allocation weights computed from the filtered covariance matrix. Thus, ideally, the filtered covariance matrix should contain as much of the persistent structure as possible. The nature of the most likely persistent structure is of course unknown from the calibration window only. 
Figure \ref{fig:overlap} shows that there are indeed strongly persistent dependence structures of asset price returns between two non-overlapping periods. Similarly to correlation matrices of DNA microarray data, while a pure HC does capture a sizeable part of the useful structure, the non-diagonal correlation patterns blocks e.g., around $(x,y)=(140,600)$ indicate that HC itself is not sufficient. 

Here, we propose a method that improves on hierarchical clustering. We exploit the fact that the less adequate a hierarchical ansatz, the more fragile it is with respect to small data perturbations. At a global level, the idea is thus to take bootstraps of the data and to average the resulting hierarchical structures. More precisely, we apply HCAL to bootstraps of the original data and then average all HCAL-filtered matrices to obtain a new kind of filtered matrix. We call our method BAHC, which stands for Bootstrapped Average Hierarchical Clustering, and define it for covariance and correlation matrices. BAHC rests on multiple hierarchical structures weighted by their frequency. A single hierarchical structure will only emerge if all the bootstrap realizations lead to the same dendrogram. Thus, this method is particularly adapted to data that is well-described by a hierarchical structure in a first approximation~\cite{mantegna1999hierarchical} but avoids selecting a single fragile structure.

\section*{Results}

\subsection*{Microarray DNA}

\begin{figure*}
\centering
\label{fig:luscgenesamp_proj}\includegraphics[width=1.9\columnwidth]{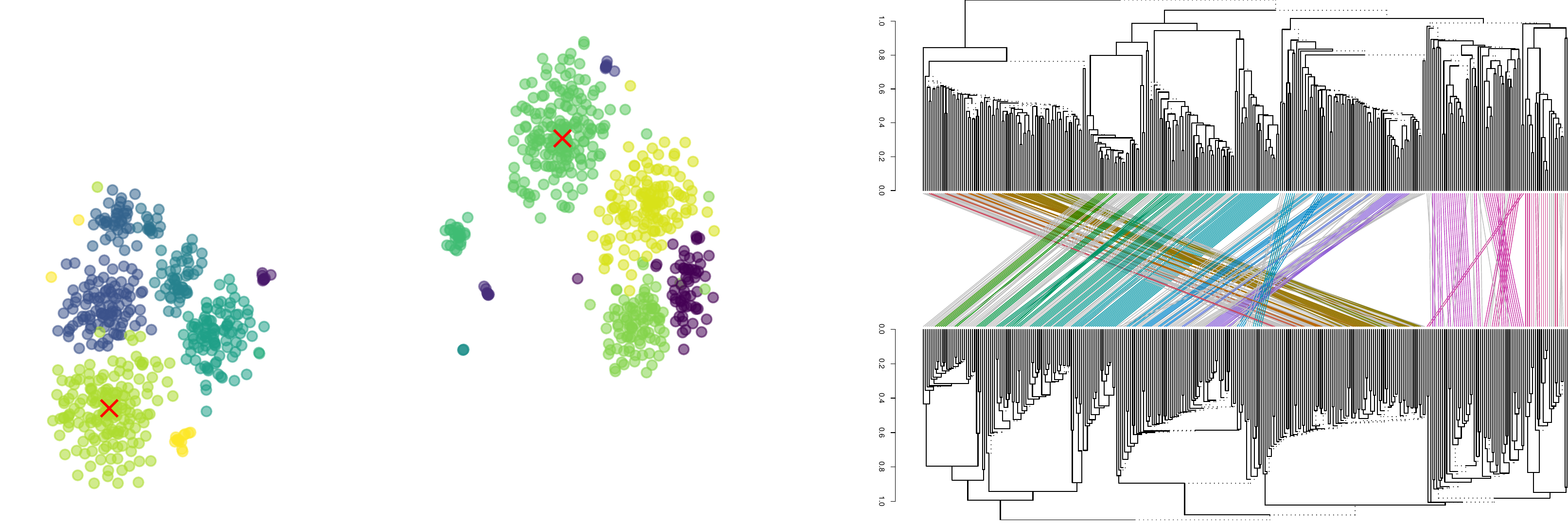}
\caption{Bidimensional t-SNE projection of the cophenetic distance between the dendrograms of 1000 bootstraps of DNA microarray data \cite{yeoh2002classification}. Two main clusters emerge, with further subclusters, corresponding to distinct potential hierarchies of dependence that are compatible with data. The red crosses indicate the centroids of the two largest clusters whose structure differences appear in the tanglegram of right plot.}\label{fig:DNAproj}
\end{figure*}

We first apply the BAHC method to DNA microarray data~\cite{yeoh2002classification} where the objects are $n=327$ tissues of patients affected by pediatric acute lymphoblastic leukemia and features are the expression intensities of $t=271$ genes ($q\simeq 1.21$). Classifying leukemia subtypes based on their gene expression profile is crucial to correct prognosis and risk assessment. However, the simplistic classification obtained from a single tree could lose relevant information coming from the complex interactions among the elements analyzed.

To show the new insights brought by BAHC compared to a simple hierarchical clustering, we kept the dendrograms of all the bootstraps and produced a bidimensional t-SNE projection \cite{maaten2008visualizing} of their cophenetic correlation coefficients. Two main clusters appear, which essentially differ by the topmost branches, as shown by the tanglegram (right plot of Fig.\ \ref{fig:DNAproj}). This means that two parts of the dendrogram which appear to be far away in a dendrogram may be much closer in another one. We applied spectral clustering \cite{ng2002spectral} to determine sub-clusters of each main cluster. Typically, sub-clusters within either of the main clusters differ at lower levels of branching. In summary, sub-groups of cancers that lie on far branches of the sample dendrograms could be miss-classified as uncorrelated despite being possibly much closer in the dendrograms of many bootstraps.

\subsection*{Risk minimization}
Given the $n\times (t+1)$ matrix of values of choice $i$ at time $k$, $p_{i,k}$, and the value returns $r_{i,k}=p_{i,k}/p_{i,k-1}-1$, one must determine the fraction of investment given to each choice $i$, the $i-$th component of vector $\mathbf{w}$. The risk is measured by the standard deviation of the portfolio return, denoted by $v_P$, whit $v_P^2=\mathbf{w}^T\Sigma\mathbf{w}$, where $\Sigma$ is the $n\times n$ covariance matrix of the matrix of returns $R$. If the weights can be negative, the optimal weights $ \tilde{\mathbf{w}} = \frac{\Sigma^{-1} \cdot \mathbbm{1}}{\mathbbm{1}^T \cdot \Sigma^{-1} \cdot \mathbbm{1}}$,  with the condition $\sum_iw_i=1$ in order to avoid the trivial solution $\mathbf{w}=0$. This situation is called long-short portfolio in the following.
In some situations, e.g., when choosing one's portfolio of energies or products, only positive weights are allowed, in which case one has to solve a quadratic programming problem; we refer to this situation as long-only portfolio.

The realized (out-of-sample) risk is the relevant performance measure. Using the  $^{out}$ exponent, the realized risk is 
$$
v_P^{out}=\sqrt{(\tilde{\mathbf{w}})^\dagger\Sigma^{out}\tilde{\mathbf{w}}},
$$
where $\tilde{\mathbf{w}}$ are computed from the in-sample covariance matrix, filtered or not, and $X^\dagger$ is the transpose of matrix $X$.

All the results reported below use the simulation setup described in the Methods section: in short, we perform 10,000 simulations of $n=100$ random assets in random periods.
We compare the out-of-sample risk computed from BAHC and several other well-known methods: the classic Ledoit and Wolf linear shrinkage method (LW henceforth) \cite{ledoit2004well} and the more recent nonlinear shrinkage approach based on the inversion of the QuEST function (QuEST)~\cite{ledoit2012nonlinear}. We also include the Cross-Validated eigenvalue shrinkage (CV)~\cite{bartz2016cross} and HCAL \cite{tumminello2007hierarchically}, denoted by $<$.

\begin{figure}
\centering
\includegraphics[width=0.45\columnwidth]{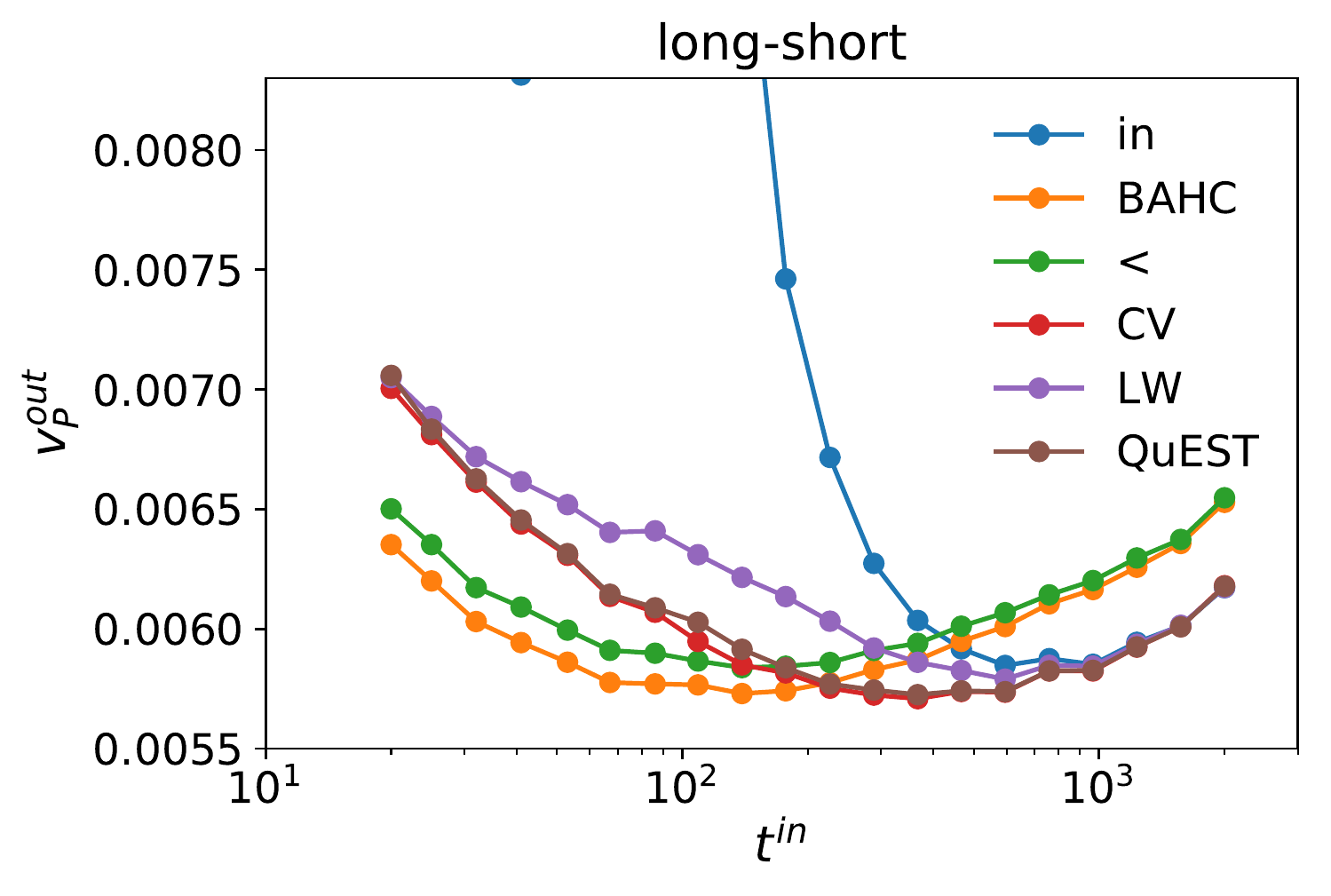}\includegraphics[width=0.45\columnwidth]{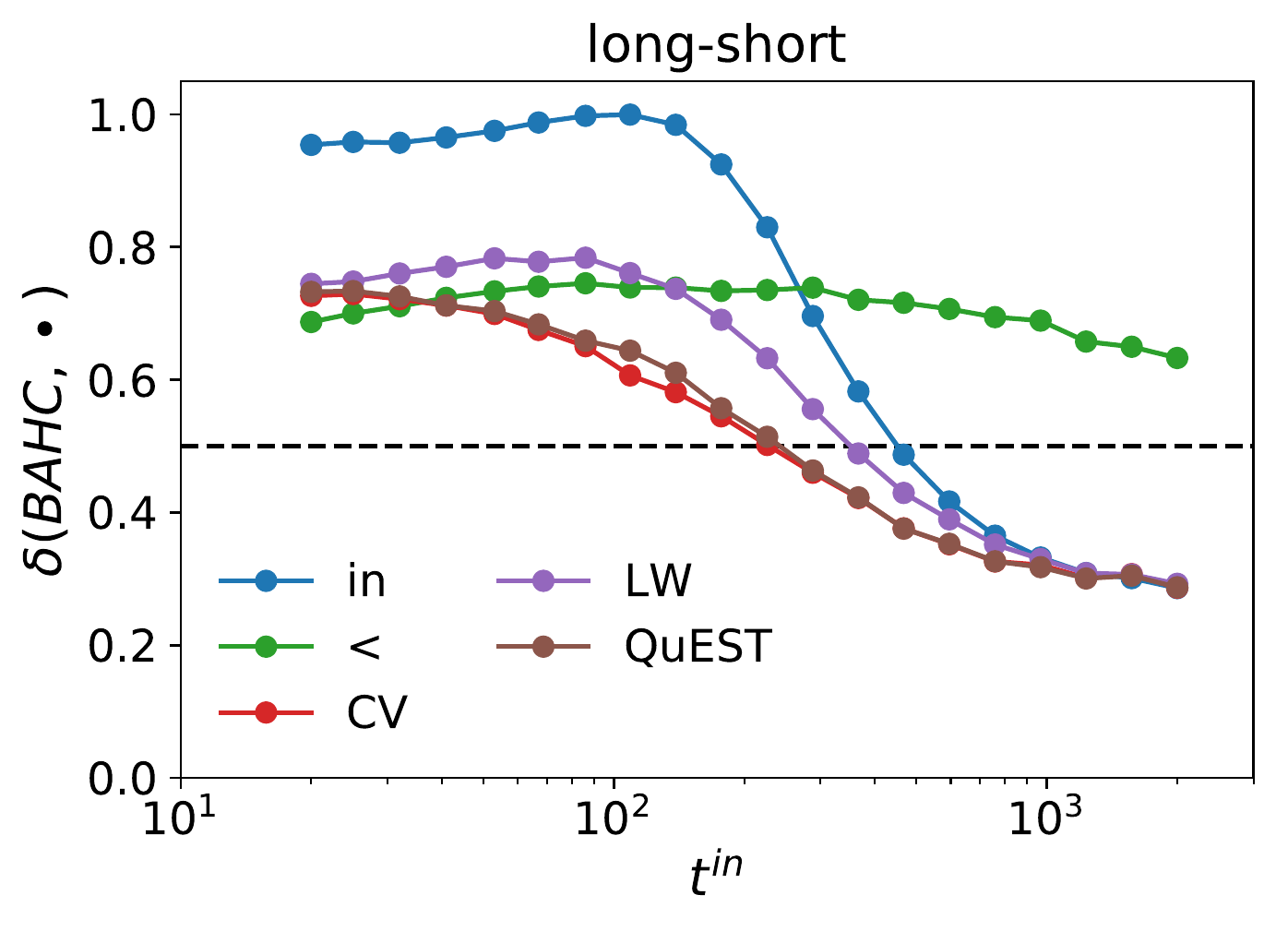}

\includegraphics[width=0.45\columnwidth]{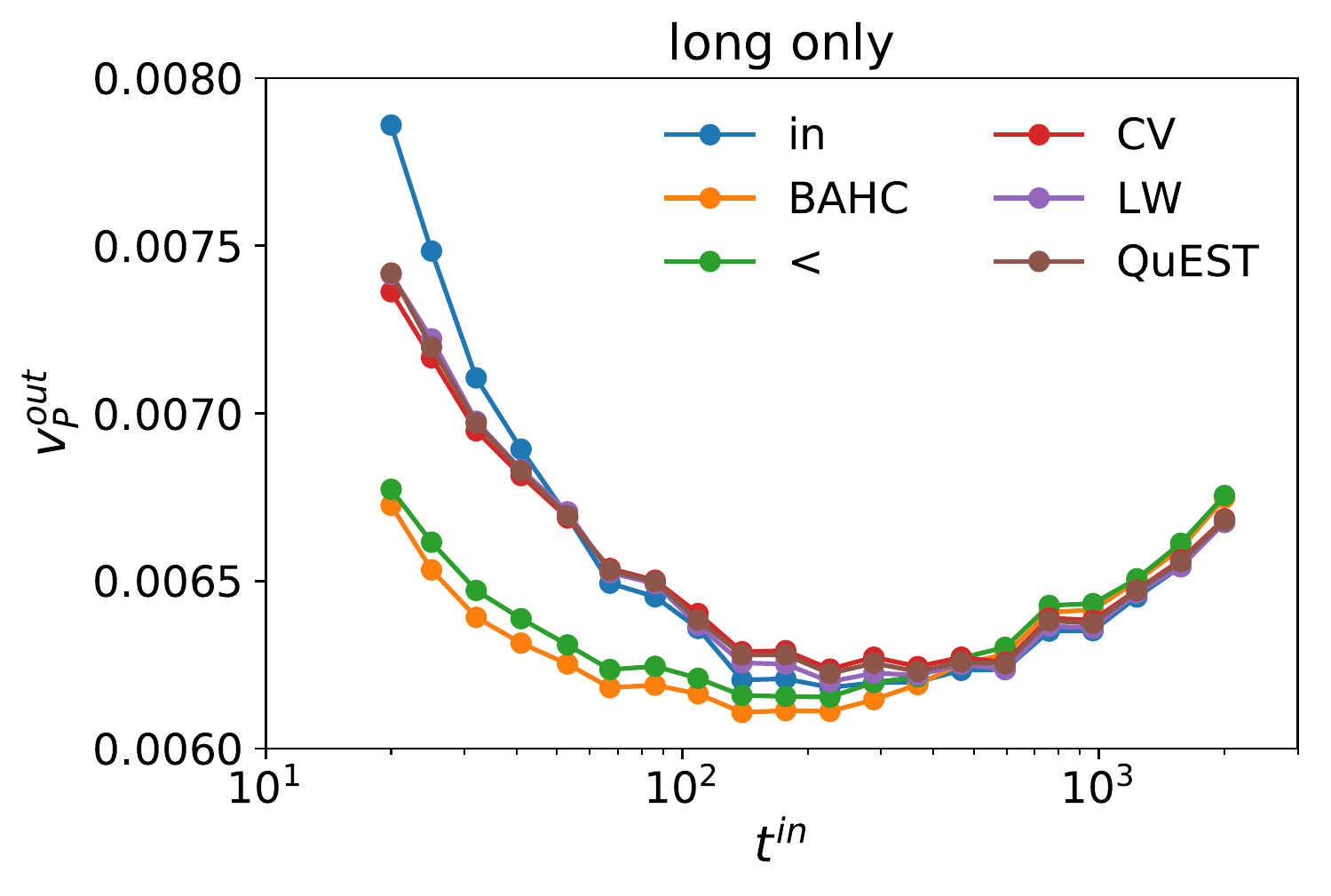}\includegraphics[width=0.45\columnwidth]{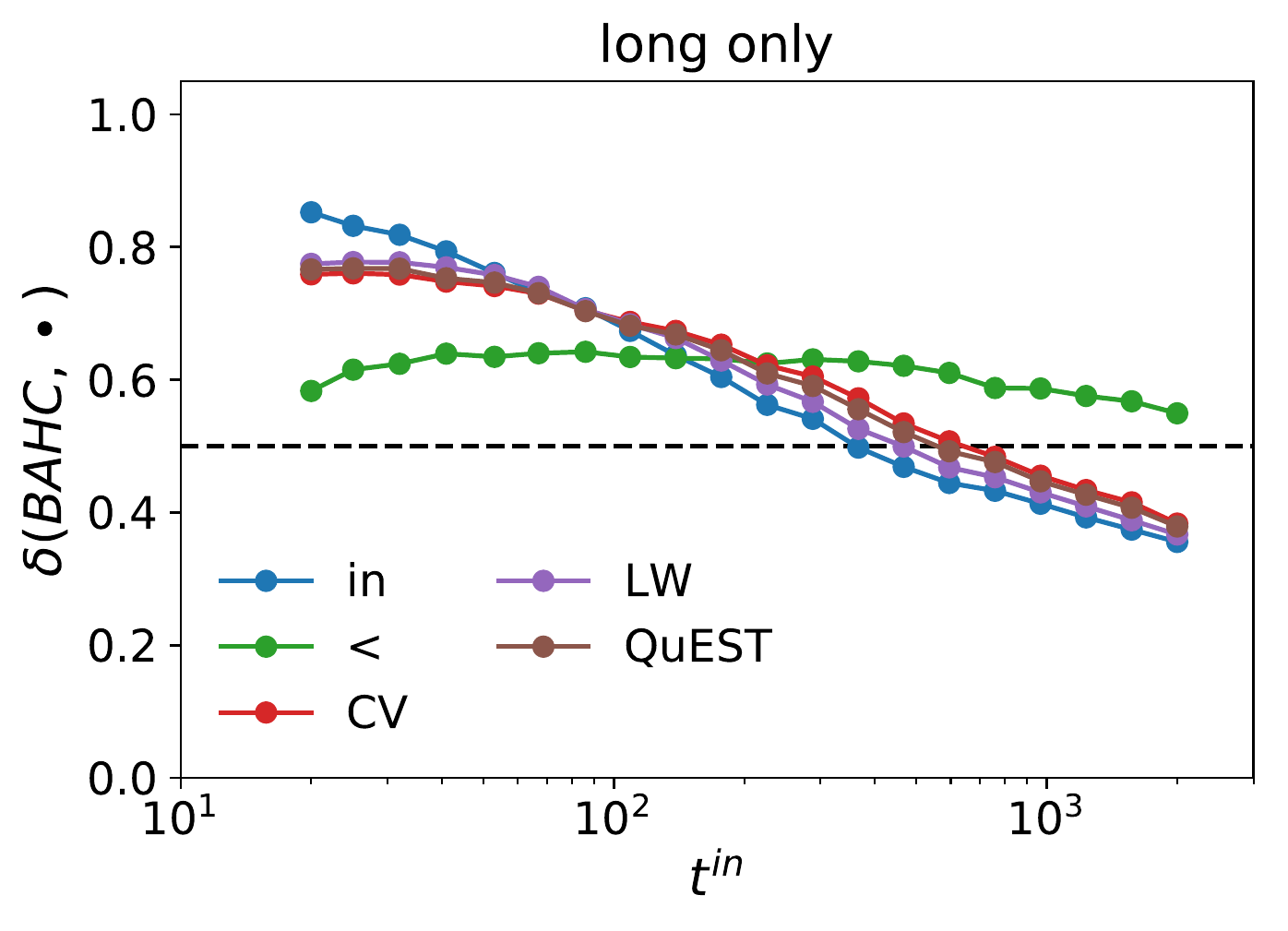}
\caption{Left plots:  realized risk for different estimators; right plots:  fraction of time the realized risk of BAHC is smaller than the one obtained with alternative estimators.  $10,000$ independent simulations per point; $t^{out}=42$ days, $n=100$ assets, US equities.\label{fig:ptf}}
\end{figure}

Figure~\ref{fig:ptf} shows that BAHC outperforms all the alternative methods  for $t^{in}\lesssim 300$, i.e., for $q=n/t \gtrsim \frac{1}{3}$, which includes all of the high-dimensional regime $q>1$. 
In particular, for the long-only portfolios, the BAHC method reaches the absolute minimum out-of-sample risk over all $t^{in}$ and all methods for $t^{in}\simeq 200$, i.e., $q\simeq 1/2$. The right-hand-side plots of Fig.~\ref{fig:ptf} report the probability that BAHC outperforms each alternative method when $q> 1/2$, which confirms that BAHC is better than all the other methods not only with respect to the average realized risk, but also in probability in this region.

\begin{figure}
\includegraphics[width=0.45\columnwidth]{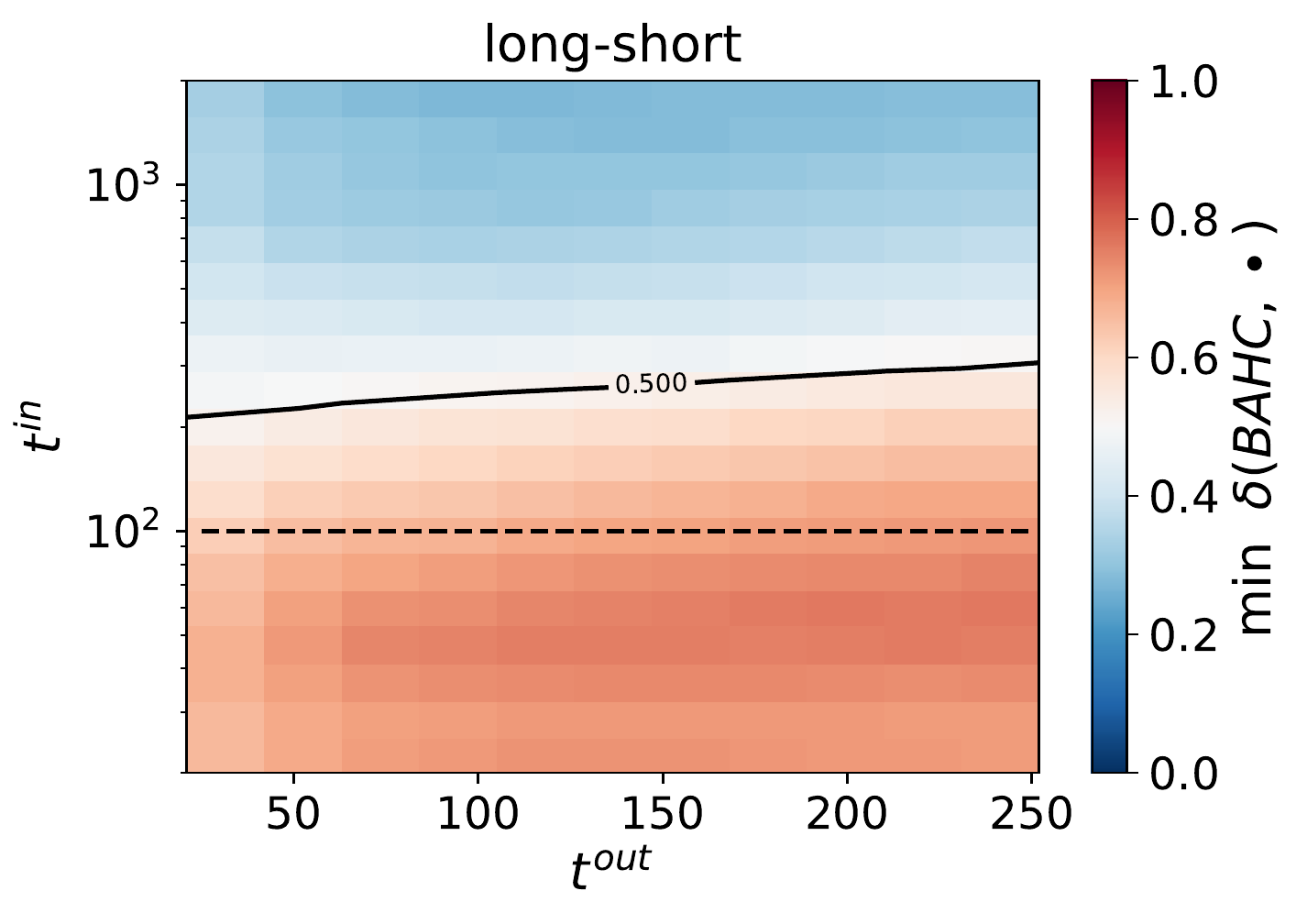}
\includegraphics[width=0.45\columnwidth]{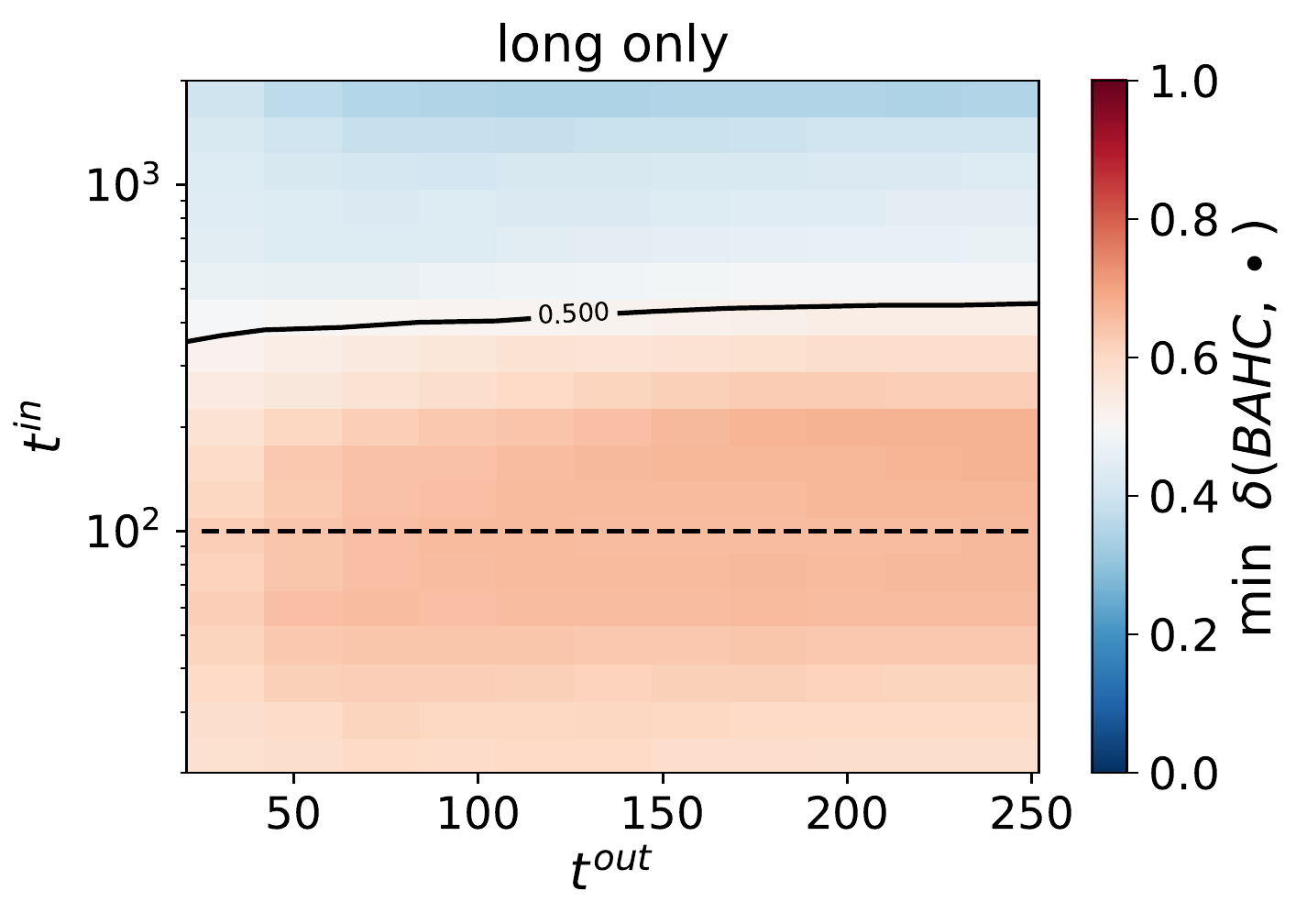}
\caption{Fraction of time BAHC yields a smaller realized risk than all the alternative methods.  Left plot: portfolios with positive and negative weights; right plot: portfolios with only positive weights. The dotted line corresponds to $q=t/n=1$, and the level curve to a 50\% probability. $10,000$ independent simulations per point; $t^{out}=42$ days, $n=100$ assets, US equities.\label{fig:fptfout}}
\end{figure}

Finally, we vary the length of the test window, $t^{out}$. We report the probability that the BAHC method outperforms all its competitors as a function of both $t^{in}$ and $t^{out}$ in Fig.~\ref{fig:fptfout}. Our approach achieves lower realized riskwith in more than half the simulations than any other method tested here as soon as $t^{in}<226$ ($q>1/2.26$) for every $t^{out}$ in the considered range. Remarkably, as $t^{out}$ increases, the calibration length below which BAHC has better than 50\% chances to outperform all its competitors only weakly increases. We interpret this result by the fact that our method is able to extract the right kind of persistent structure in that particular data, which is confirmed below by spectral analysis. We found similar results for the Hong Kong equity market (see S.I.).

\subsection*{Spectral Properties}

In order to understand why and when our method has a better performance than the other methods based on spectral clustering, it is instructive to compare the in- and out-of-sample persistence of the eigenvalues and eigenvectors produced by all the filtering methods considered here. The spectral decomposition of correlation matrix $C$  is denoted by $C = U^\dagger \Lambda U$, where $U$ is a $n\times n$ matrix formed by the eigenvectors of $C$ and $\Lambda$ is the diagonal matrix obtained from the corresponding eigenvalues. 

\subsubsection*{Eigenvectors stability}

A simple way to characterise eigenvectors stability is to compare the empirical out-of-sample correlation matrix $C^{out}$ with the Oracle correlation estimator defined as $\Xi^{in}_C={U^{in}}^\dagger Z^{in}U^{in}$ where $Z^{in}=\textrm{diag }({U^{in}}^\dagger C^{out}U^{in})$ is the Oracle eigenvector estimator, the idea being that $\Xi^{in}_C=C^{out}$ if in- and out-of-sample eigenvectors coincide (see S.I.). The Oracle estimator for the covariance matrix, denoted by $\Xi^{in}_\Sigma$, is defined in a similar way. 

\begin{figure}
\includegraphics[width=0.45\columnwidth]{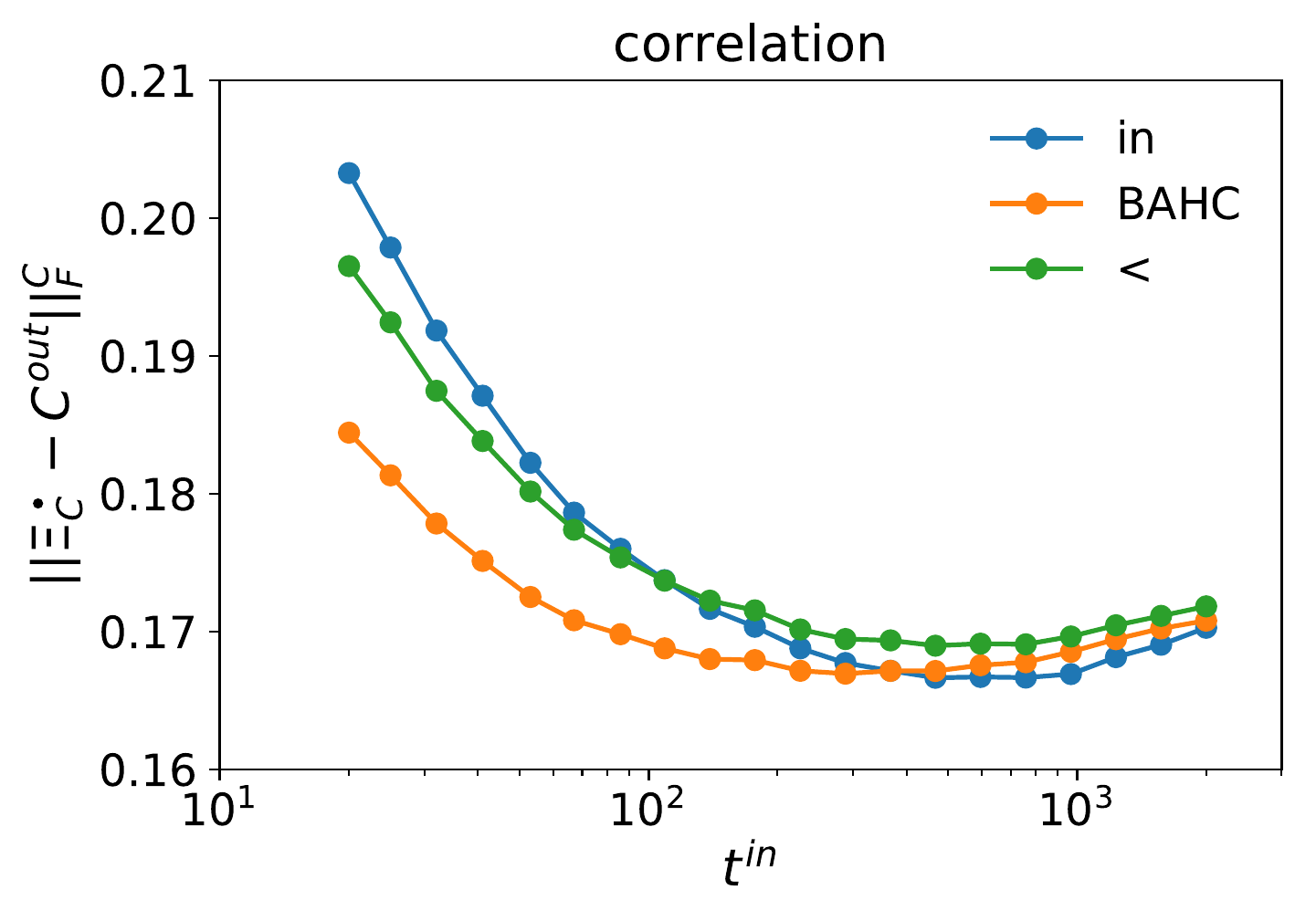}
\includegraphics[width=0.45\columnwidth]{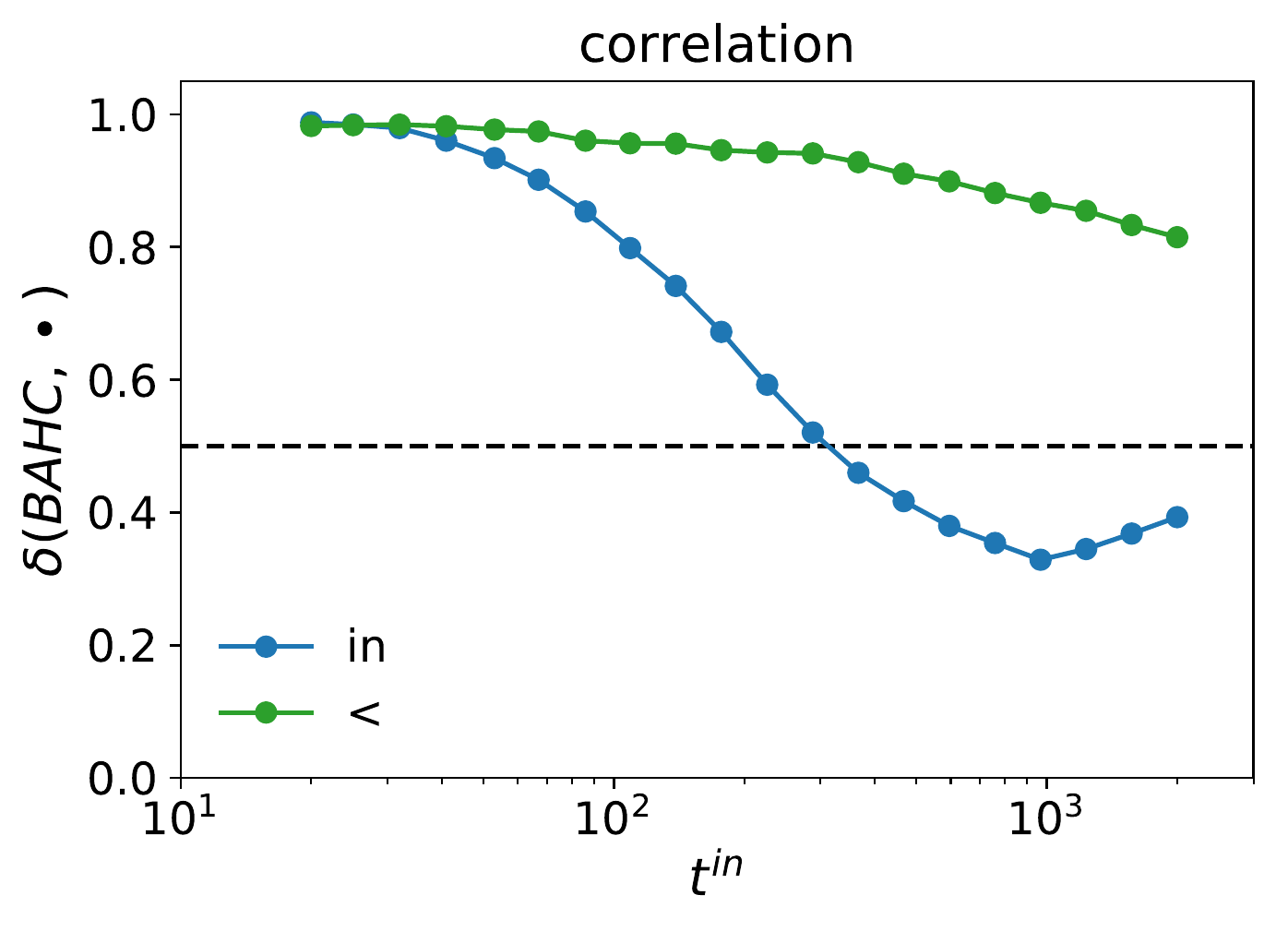}

\includegraphics[width=0.45\columnwidth]{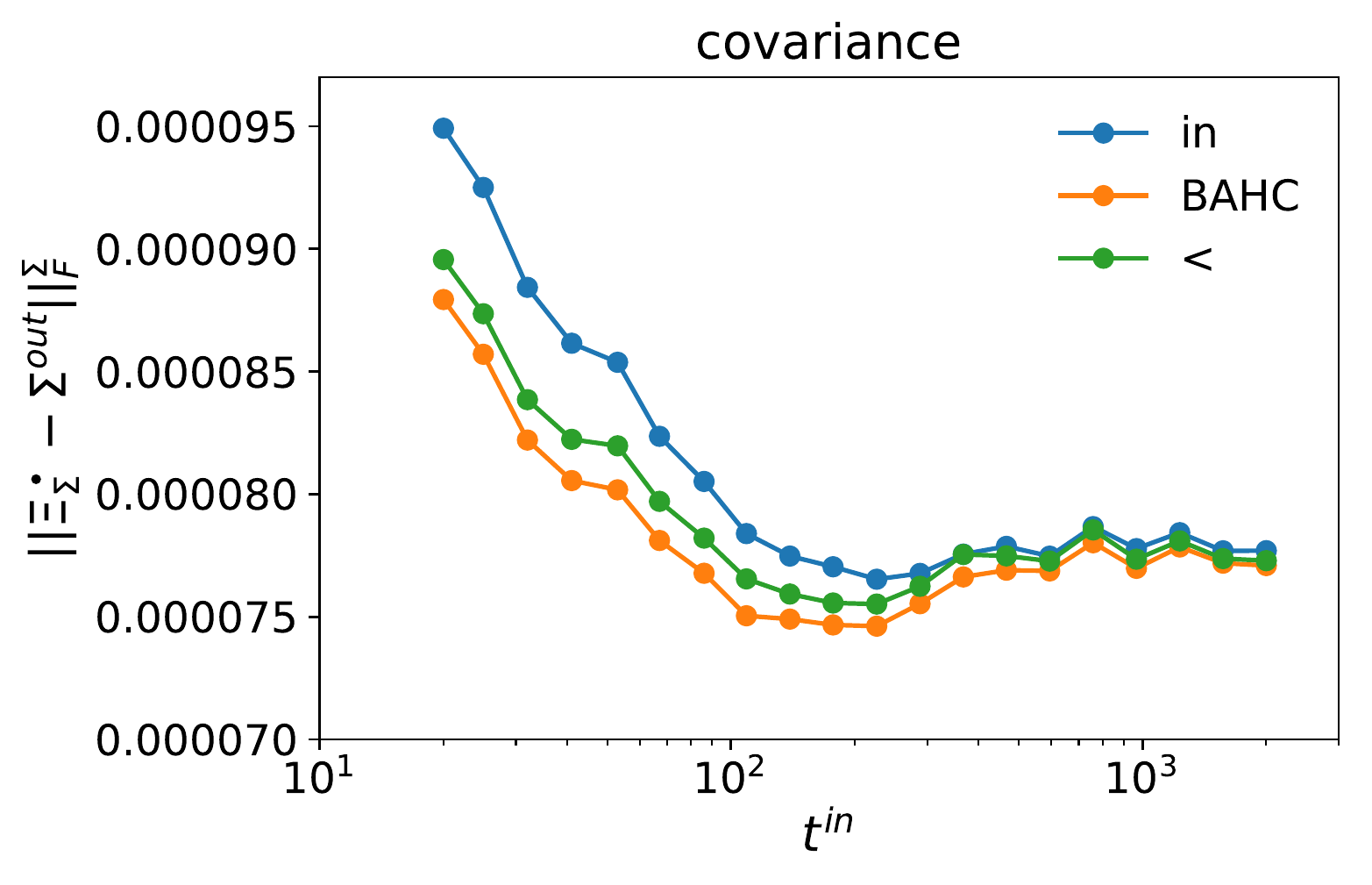}
\includegraphics[width=0.45\columnwidth]{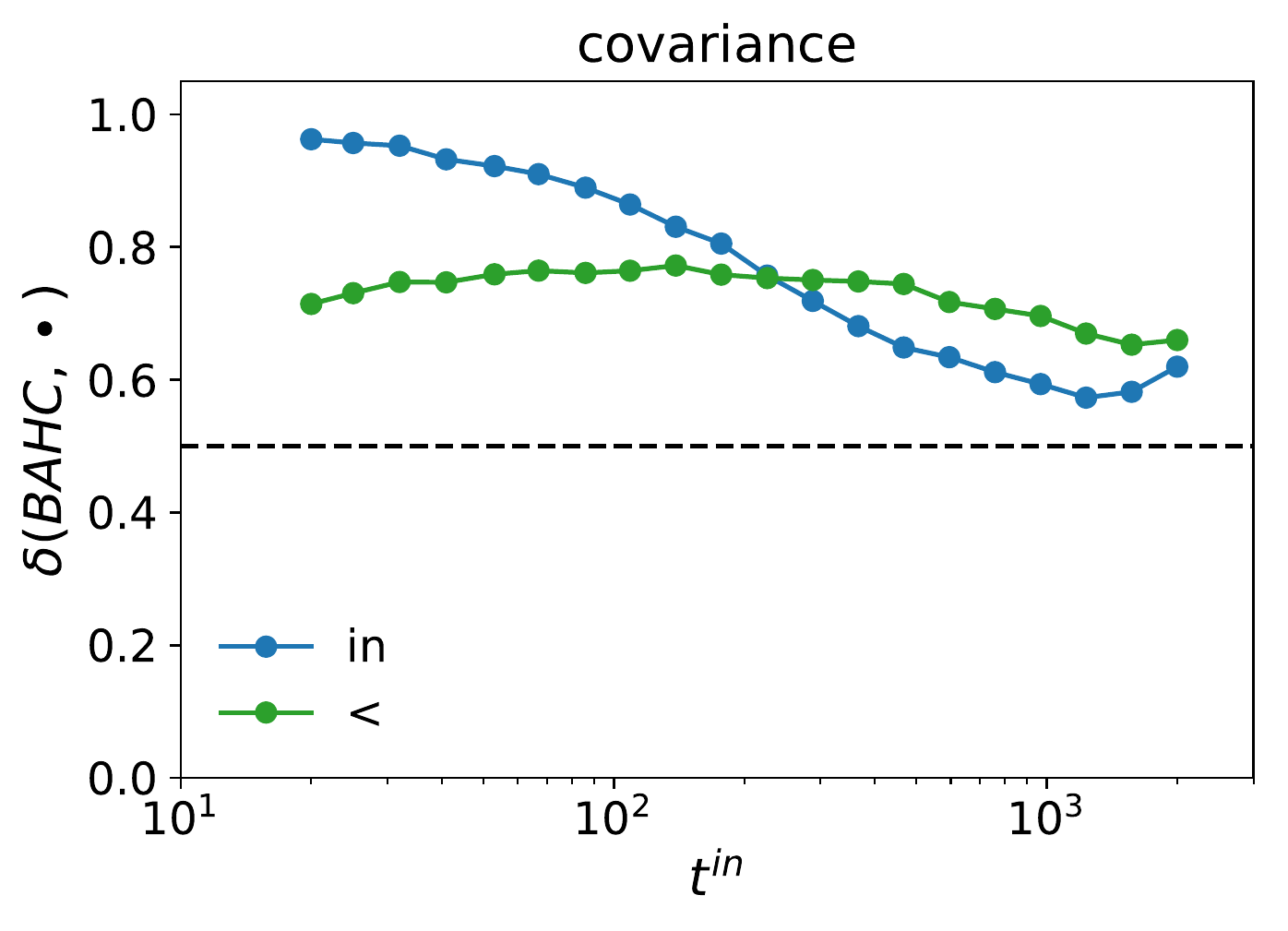}

\caption{Frobenius distance between the out-of-sample matrices and the Oracle estimators obtained with the in-sample eigenvectors ($in$), the in-sample BAHC-filtered eigenvectors ($BAHC$) and the in-sample HCAL-filtered eigenvectors ($<$). Upper panels refer to correlation matrices $C$, lower panels to covariance matrices $\Sigma$. The left panels are the Frobenius norm of the difference between the estimator and the out-of-sample realization; the right panels are the fraction of time BAHC outperforms the alternative estimators. $10,000$ independent simulations per point; $t^{out}=42$ days, $n=100$ assets, US equities.\label{fig:OracleOver}}
\end{figure}

Figure \ref{fig:OracleOver} reports the Frobenius distances (see the Methods section) $\left\lVert C^{out}-\Xi^{in}_C \right\rVert _F^C $ and $\left\lVert \Sigma^{out}-\Xi^{in}_\Sigma \right\rVert _F^\Sigma $ as a function of $t^{in}$ for $n=100$ assets. Note that CV, LW and QuEST methods all use the in-sample eigenvectors and thus do not need separate computations. Generally, our method yields more stable correlation and covariance matrices for $t^{in}<300$ ($q>1/3$), i.e., already in the low-dimensional case.  The difference is due to the fact that the eigenvectors obtained by our method are more stable than the vanilla in-sample eigenvectors, which mechanically improves the Oracle estimator. 

Figure~\ref{fig:OracleOver} also shows that the probability that the eigenvectors of BAHC-filtered correlation matrices are more stable than those provided by the alternative filtering methods grows as $t^{in}$ becomes smaller. The same applies to the comparison between BAHC -filtered and empirical covariance matrices, while HCAL, denoted by $<$, has better performance in about a 25\% of samples almost independently of $t^{in}$. In short, as soon as $q>1/3$ in this dataset, the BAHC method likely yields more persistent eigenvectors than all the other filtering methods considered here.

\subsubsection*{Eigenvalues stability}

Since both the covariance $\Sigma$ and precision $\Sigma^{-1}$ matrices are relevant to minimum-variance optimization, we measure two types of residues that focus on large and small eigenvalues, defined as
\begin{eqnarray}
\epsilon_{hi} = \sqrt{\frac{1}{n}\sum_{i=1}^n\left( \lambda_{i}-z_{i}\right)^2	}\label{eq:rehi}\\
\epsilon_{low} = \sqrt{\frac{1}{n}\sum_{i=1}^n\left( \frac{1}{\lambda_{i}}-\frac{1}{z_{i}}\right)^2,	} \label{eq:relow}
\end{eqnarray}
where $\lambda_{i}=(\Lambda)_{ii}$ is the $i$-th eigenvalue of the in-sample estimator and $z_{i}=(Z^{in})_{ii}$ comes from the Oracle estimator computed with the respective filtered eigenvector matrix and $i$ is the respective rank of these eigenvalues. The residue measure $\epsilon_{hi}$ mainly  accounts for the discrepancy between the largest eigenvalues and the residue measure $\epsilon_{low}$ attributes more weight to the discrepancy between the smallest eigenvalues.

\begin{figure}
\includegraphics[width=0.45\columnwidth]{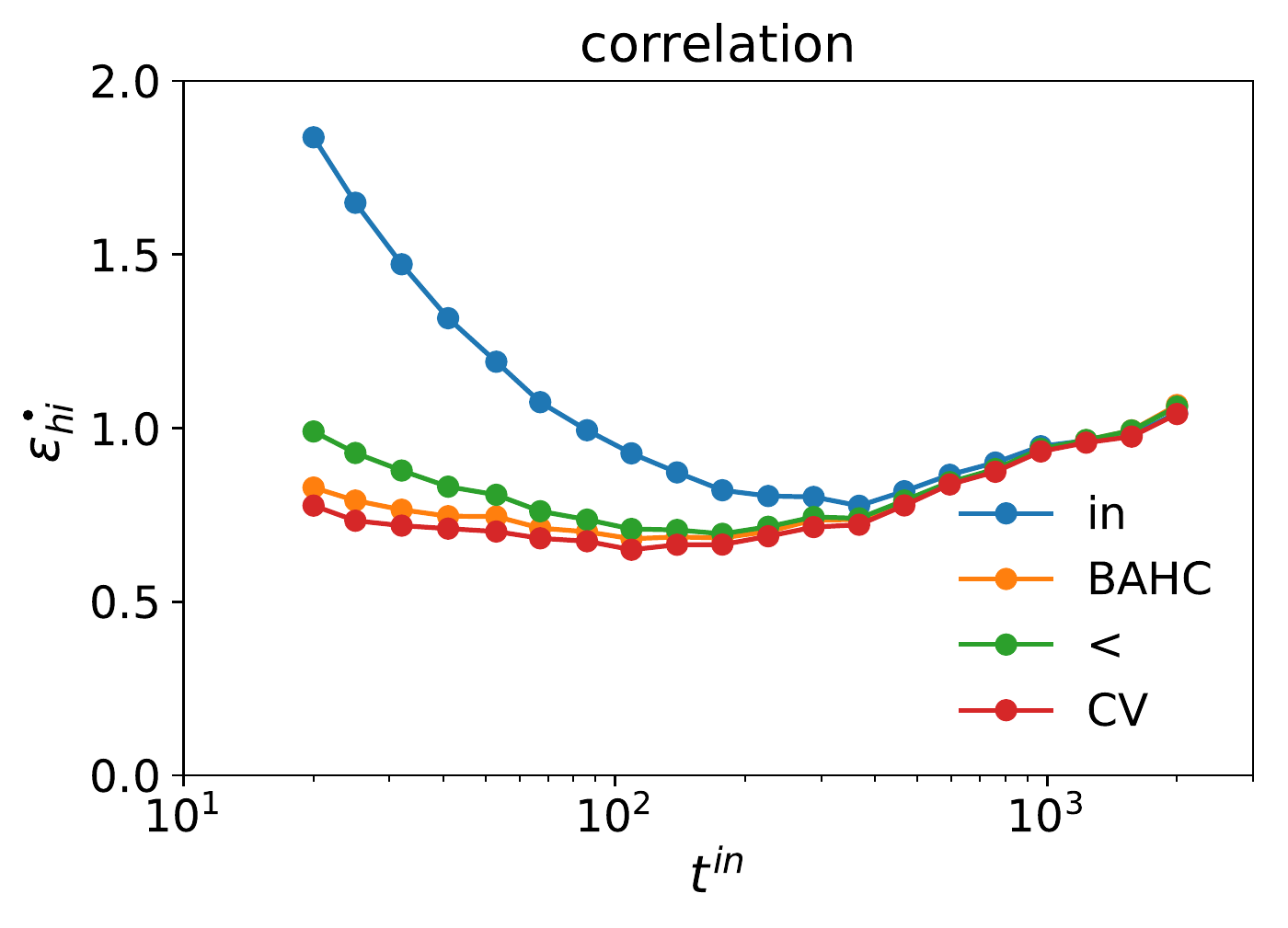}
\includegraphics[width=0.45\columnwidth]{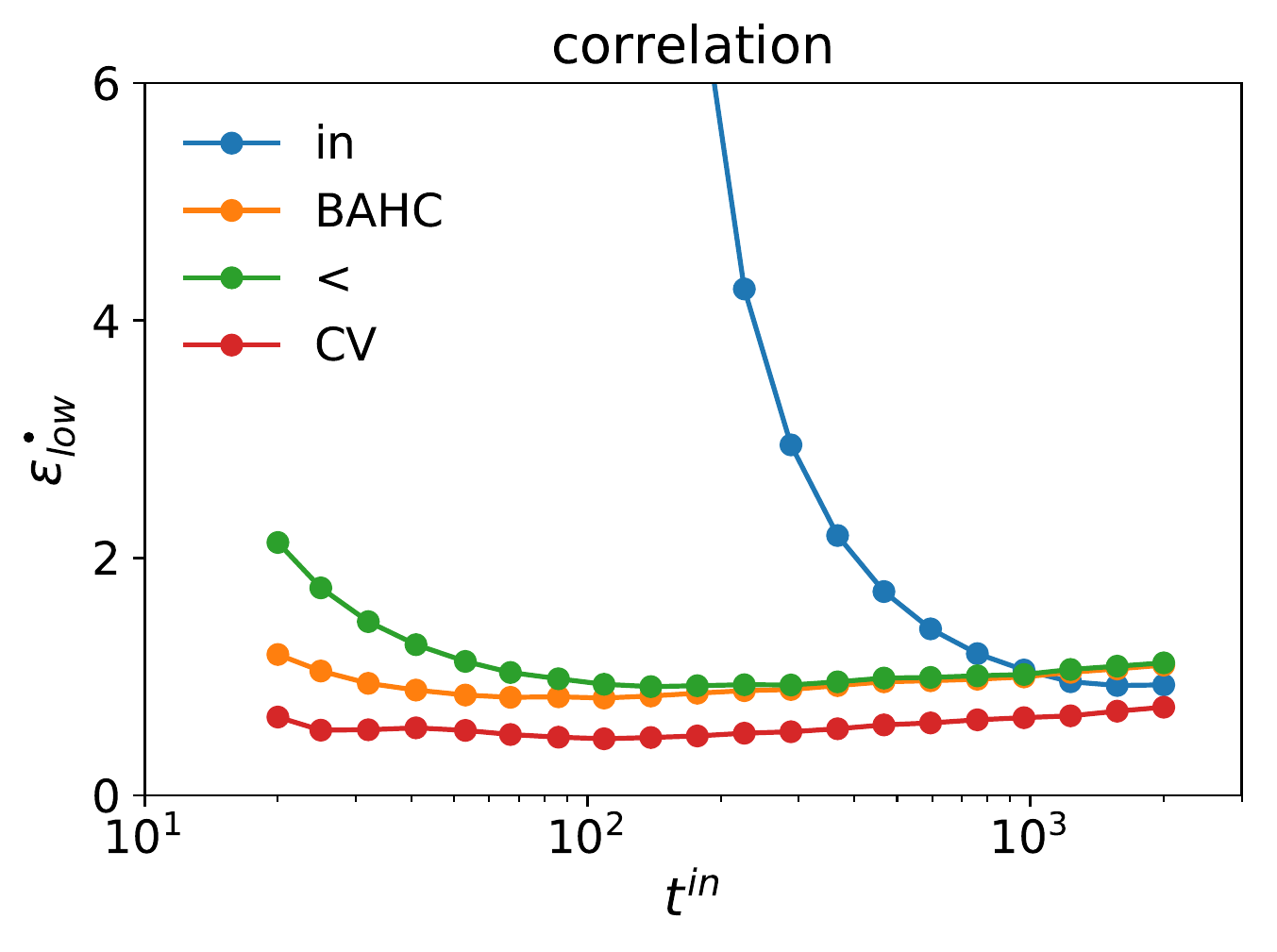}

\includegraphics[width=0.45\columnwidth]{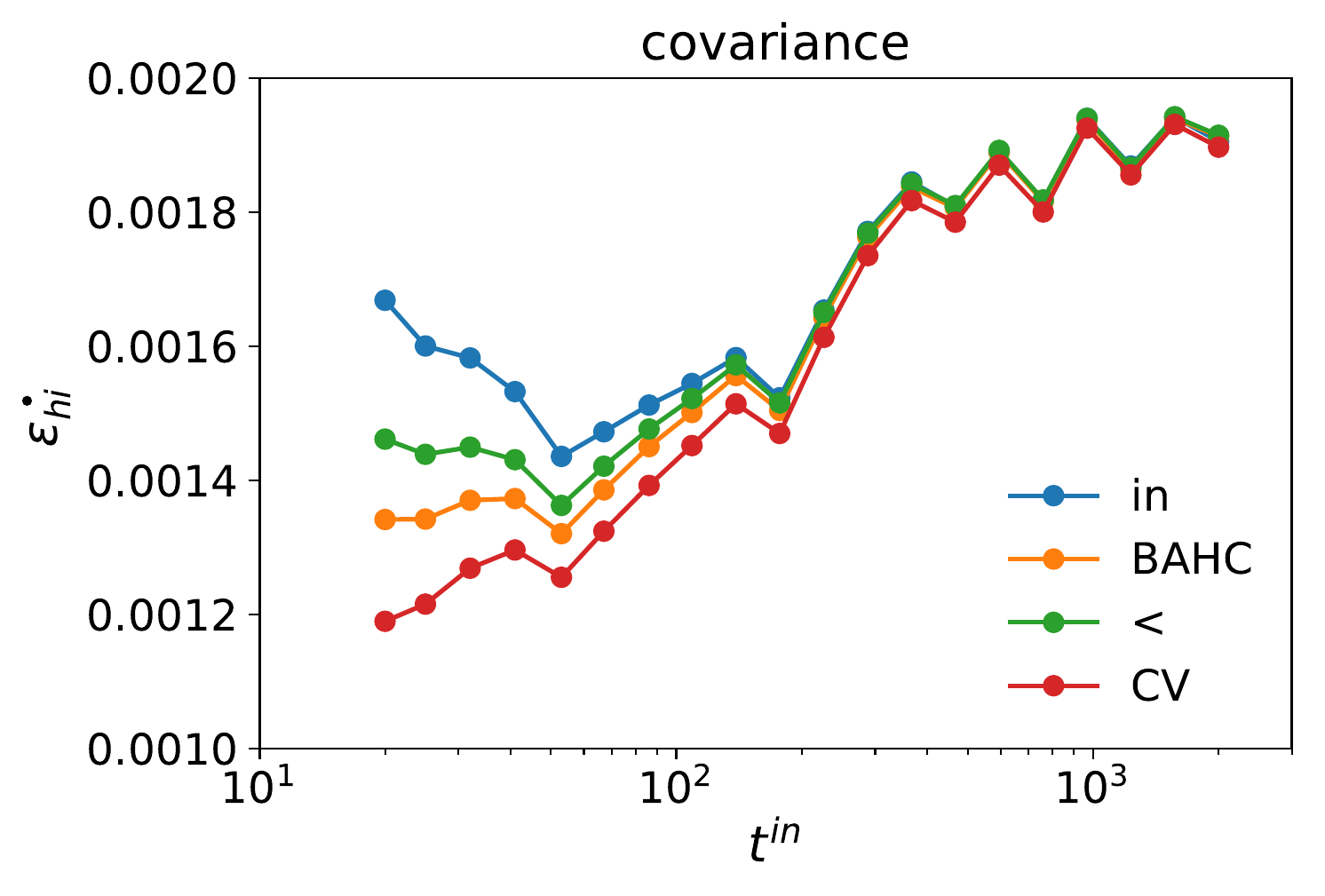}
\includegraphics[width=0.45\columnwidth]{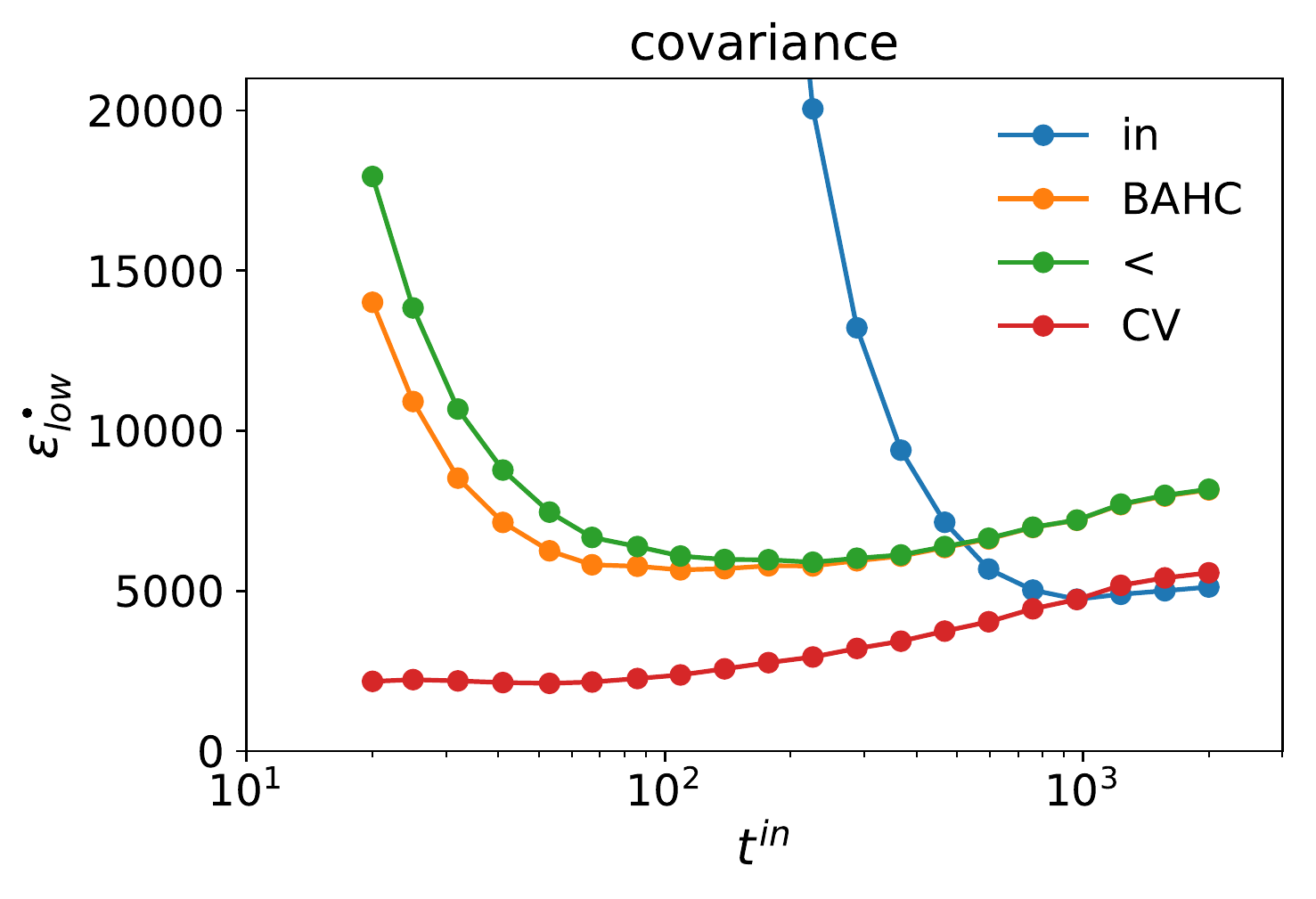}
\caption{Average residue  $\epsilon_{hi}$ and $\epsilon_{low}$ over $10,000$ simulations with random calibration windows and a random selection of $n=100$ assets. The upper panel refers to the correlation matrix, the lower panel refers to the covariance matrix. $10,000$ independent simulations per point; $t^{out}=42$ days, $n=100$ assets, US equities.}\label{fig:OracleEig}
\end{figure}

Figure~\ref{fig:OracleEig}  plots the residues of the correlation and covariance matrices respectively as a function of $t^{in}$. We compare our approach with the sample estimator, HCAL-filtered matrix, and the Cross-Validated (CV) eigenvalue distribution. While  CV method outperforms all the other methods when $t^{in}\lesssim 1000$ ($q>0.01$), the eigenvalues produced by our method are still much closer to the Oracle than those of the raw sample estimator when $t^{in}\lesssim 500$.

\subsection*{Filtered correlation and covariance matrices}

\begin{figure}
\includegraphics[width=0.45\columnwidth]{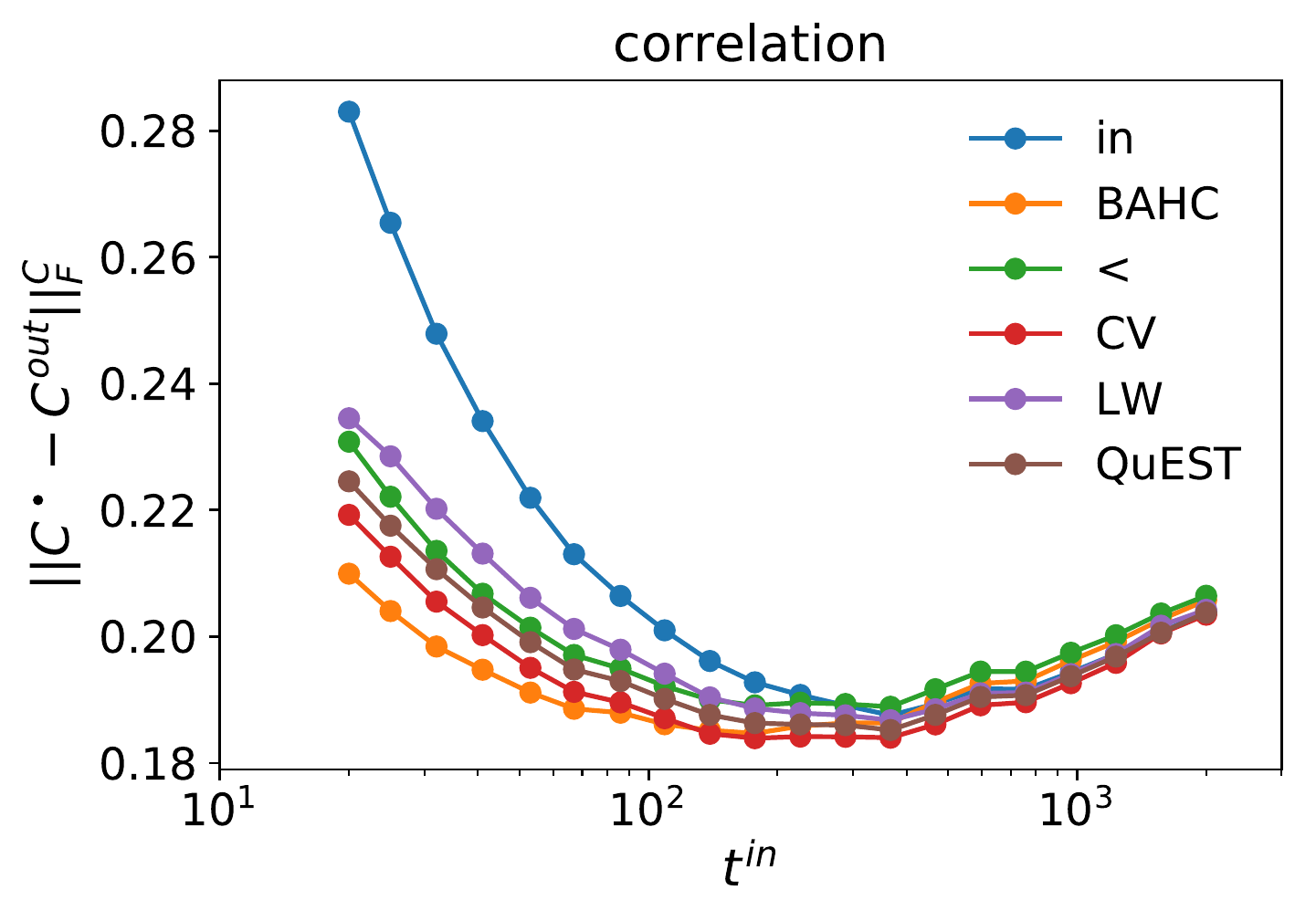}\includegraphics[width=0.45\columnwidth]{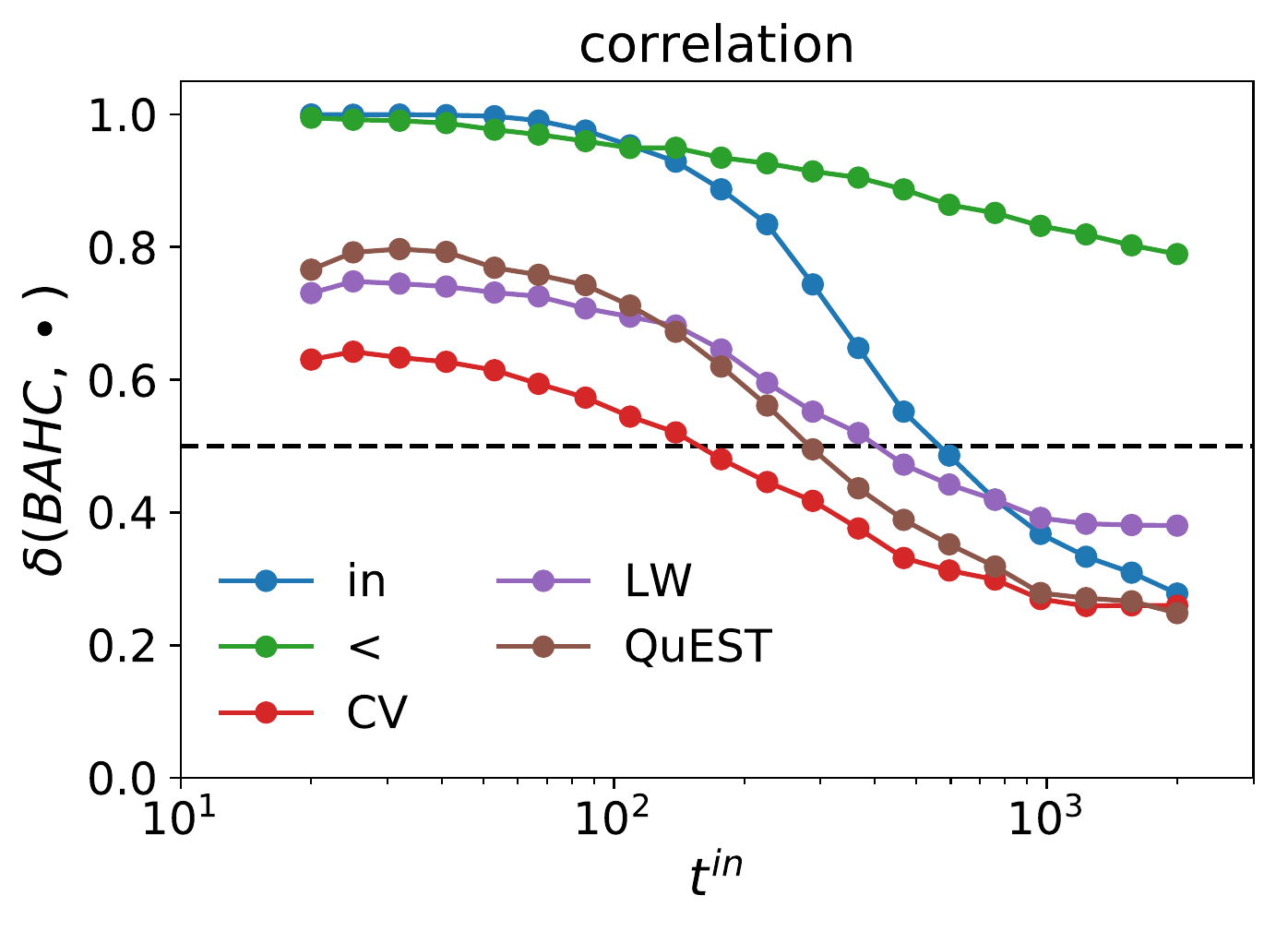}
\includegraphics[width=0.45\columnwidth]{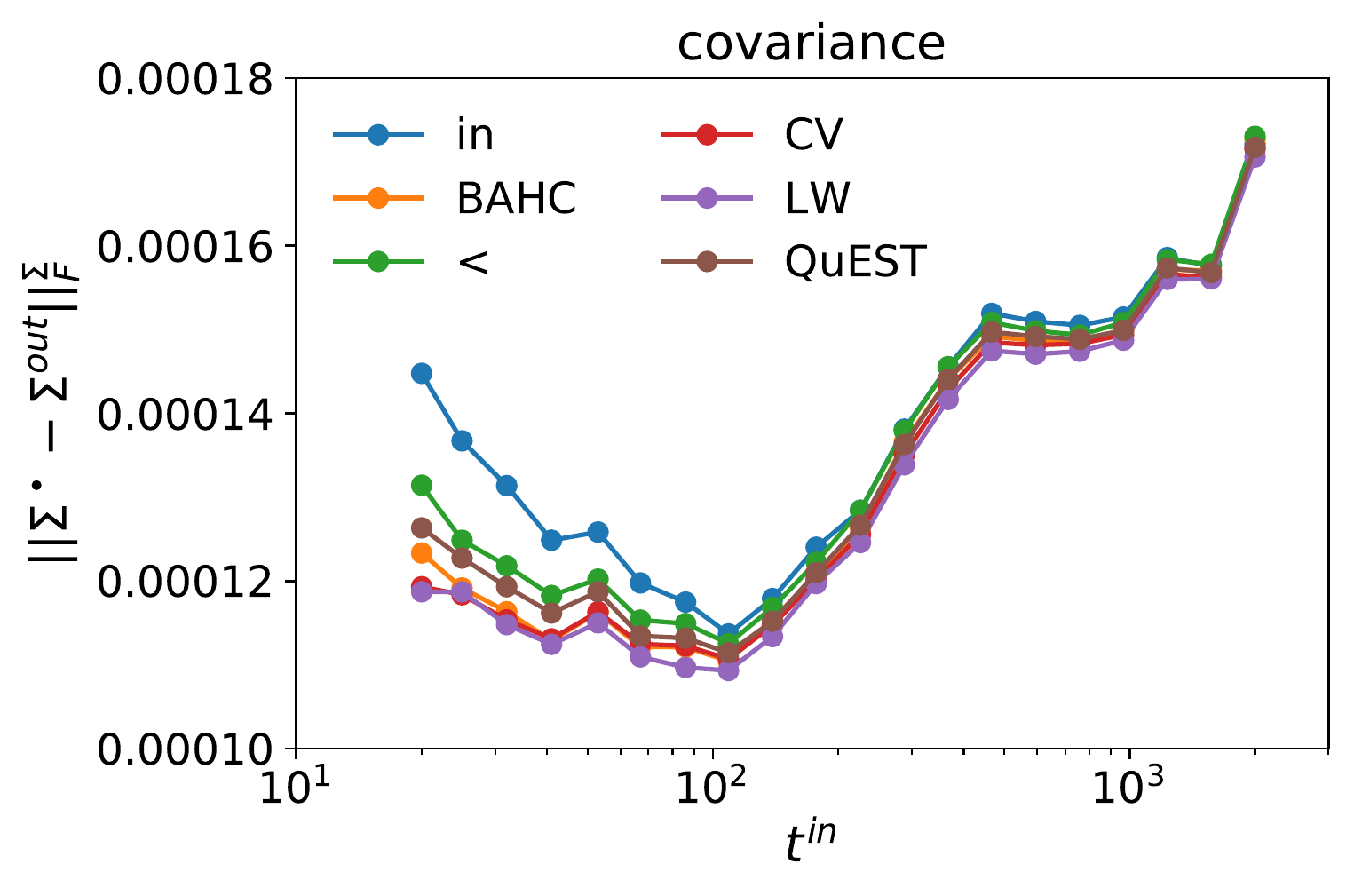}\includegraphics[width=0.45\columnwidth]{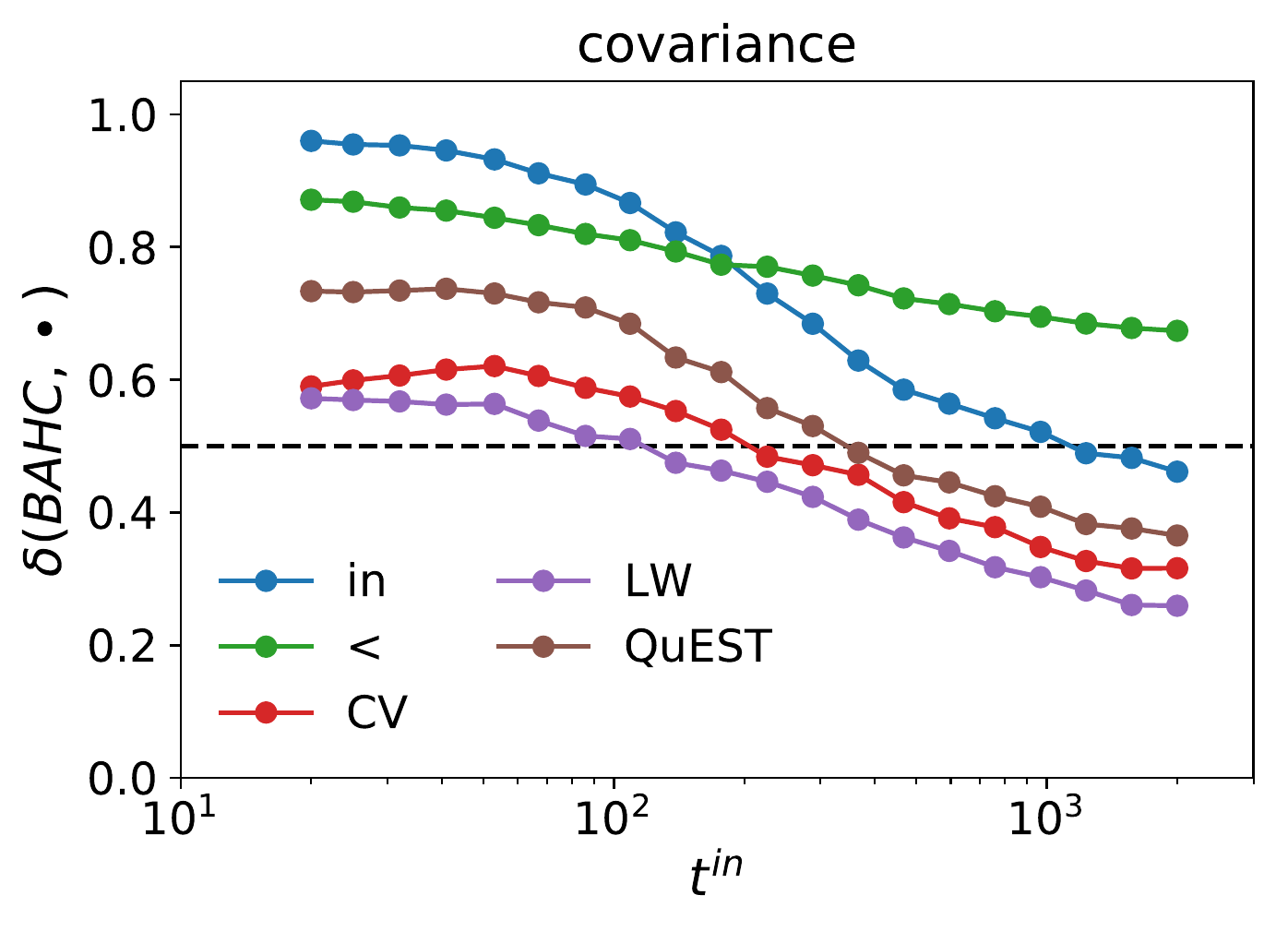}
\caption{Left plots: Frobenius distance between out-of-sample matrices and filtered in-sample matrices; upper panels refer to correlation matrices $C$, lower panels to covariance matrices $\Sigma$. Right plots: Fraction of time the Frobenius distance of BAHC-filtered matrices is smaller than the alternative estimators. $10,000$ independent simulations per point; $t^{out}=42$ days, $n=100$ assets, US equities.\label{fig:corcov_inout_frob}}
\end{figure}

The ultimate test is of course to compare filtered in-sample matrices with out-of-sample matrices. Figure \ref{fig:corcov_inout_frob} reports the Frobenius distance between the filtered in-sample and out-of-sample correlation and covariance matrices for all the tested methods. Expectedly, BAHC outperforms all the other ones  for $t^{in}\lesssim 300$. Figure~\ref{fig:corcov_inout_frob} plots the fraction of times the Frobenius norm of our method is lower than the other methods, which shows that the BAHC method outperforms HCAL filtering for every $t^{in}$.

\section{Discussion}

Filtering covariance and correlation matrices requires to take care of $O(n^2)$ coefficients. Focusing on $O(n)$ variables, for example by tweaking the eigenvalues or using a single hierarchical ansatz, works to some extend. Making further progresses requires to filter more variables, if possible while keeping an $O(n)$ ansatz. This is what the BAHC method that we introduce achieves: by using $m$ bootstraps and applying an $O(n)$ structure,  BAHC allows some additional flexibility, while keeping the overall structure simple.

Our method both filters out estimation noise and improves the stability of the eigenvectors in a dynamical context. Indeed, the spectral decomposition of BAHC-filtered correlation matrices is close to the optimal CV method with respect to the eigenvalue distribution.  Furthermore, in the dynamical context investigated here, the eigenvectors produced by our method have a higher overlap with the out-of-sample ones than the unfiltered in-sample eigenvectors for reasonably small $q=t/n$. This is why our method leads to better minimum-variance portfolios than all the competing filtering methods when the calibration window is small. In particular, if no short selling is allowed, our approach produces, on average, the lowest-risk portfolio.

Future work is needed to characterize the average dependence structure produced by BAHC better, from both theoretical and empirical points of view. In addition, BAHC may still be too strict in some cases and thus leave out valuable information, hence, further refinements of the ansatz will need to be investigated.

\section*{Materials and Methods}

\subsection*{Datasets description}
We consider the daily close-to-close returns of US equities, adjusted for dividends, splits, and other corporate events. More precisely, the dataset consists of large-capitalization stocks, from 1992-02-03 to 2018-06-29.  The number of stocks with data varies over time: it ranges from 399 in 1992-02-06 to 723 in 2018-06-29 and is roughly constant from 2008 onwards. The list of tickers is reported in S.I.

DNA microarray data~\cite{yeoh2002classification} can be downloaded from \cite{yeoh}. It consists of gene expression intensity of 327 tissues of patients affected by pediatric acute lymphoblastic leukemia and a subset of 271 genes.

\subsection*{Numerical simulations with financial data}
\label{ss:sim_setup}

All the simulations are carried out in the same way: each point of each plot is an average over $10,000$ simulations, each of which includes an in-sample window of length $t_{in}$ and an out-of-sample window of length $t^{out}=42$ days (about two trading months) unless otherwise specified; it starts from a random day uniformly chosen in the available dataset. To have meaningful in- and out-of-sample windows given the maximum $t^{in}$ considered, the first day of the out-of-sample must be after 01-01-2000;  each simulation selects $n=100$ assets at random among the assets with no missing value in both in- and out-of-sample windows.

\subsection*{BAHC algorithm}

Given matrix $R \in \mathbb{R}^{n \times t}$, our method prescribes to create a set of $m$ bootstrap (feature-wise) copies of $R$, denoted by $\{R^{(1)},\, R^{(2)},\, \cdots, R^{(m)}\}$. A single bootstrap copy of the data matrix $R^{(b)} \in \mathbb{R}^{n \times t}$ has elements $r^{(b)}_{ij} = r_{i s^{(b)}_j}$, where ${\bf{s}}^{(b)}$ is a vector of dimension $t$ obtained by random sampling with replacement of the elements of vector $\{1, 2, \cdots, t\}$. The vectors ${\bf{s}}^{(b)}$, $b=1,\cdots,m$ are independently sampled.

The Pearson correlation matrix of each bootstrapped data matrix $R^{(b)}$ is then computed and denoted by $C^{(b)}$; in turn the latter is filtered with the hierarchical clustering average linkage (HCAL) proposed in \cite{bongiorno2019nested}, which yields $C^{(b)<}$. In short, the HCAL uses two ingredients: the distance $D=1-C$ to agglomerate cluster in a hierarchical way, and the averaging of the correlation between clusters (see S.I. and \cite{bongiorno2019nested} for more details).

Finally, the filtered correlation matrix $C^{\textrm{BAHC}}$ is the average of the HCAL-filtered matrices $C^{(b)<}$
$$
C^{\textrm{BAHC}}=\frac{1}{m}\sum_{b=1}^m C^{(b)<}
$$

To build a BAHC-filtered covariance matrice, we estimate the variance of $r_i^{(b)}$, denoted by $\sigma_{ii}^{(b)}$, compute the HCAL-filtered covariance matrices $\Sigma^{(b)<}$ whose elements are defined as
\begin{equation}
\sigma_{ij}^{(b)<} = c_{ij}^{(b)<}\, \sqrt{\sigma_{ii}^{(b)}\, \sigma_{jj}^{(b)} },
\end{equation}
and finally obtain the BAHC-filtered covariance matrix
$$
\Sigma^{\textrm{BAHC}}=\frac{1}{m}\sum_{b=1}^m \Sigma^{(b)<}
$$

\subsection*{Frobenius norms}

We use rescaled Frobenius norms to account for the fact that the number of assets in our dataset depends on time:
\begin{equation}\label{eq:frobcov}
\left\lVert X \right\rVert _F^{\Sigma} = \sqrt{\sum_{i,j}^{n \times n} \frac{x_{ij}^2}{n^2}}.
\end{equation}
In addition, because CV, LW and QuEST methods do not guarantee the identity on the diagonal of filtered correlation matrices; therefore, contrarily to BAHC, we do not include the diagonal elements in the metric and thus define
\begin{equation}\label{eq:frobcor}
\left\lVert X \right\rVert _F^C = \sqrt{\sum_{i>j}^{n \times n} \frac{2\,x_{ij}^2}{n(n-1)}}.
\end{equation}
We found that the performance of CV, LW, QuEST-based correlation estimators is slightly improved by replacing $c_{ij}$ with $\frac{c_{ij}}{\sqrt{c_{ii}\, c_{jj}}}$, which also ensures that the diagonal elements equal one, and thus have used this modification in our analysis.

\subsection*{Source code}

We have written a BAHC package for both R and Python, available from CRAN and PyPI, respectively.


\bibliography{report_01}

This publication stems from a partnership between CentraleSupélec and BNP Paribas.

\newpage

\appendix
\section{Supporting Information Appendix (SI)}

\subsection{Average Linkage Filtered Correlation Matrix}\label{sec:AVCM}
\subsubsection{The Notation}
We describe in this section the strictly hierarchical method of Ref~\cite{tumminello2007hierarchically}. Given a generic matrix $R \in \mathbb{R}^{n \times t}$, a generic $\sigma_{ij}$ element of the $n \times n$ sample covariance matrix is defined as
\begin{equation}
\sigma_{ij} =\frac{1}{t} \sum_{h=1}^t \left( r_{ih} - \bar{r}_i \right) \left(r_{jh} - \bar{r}_j \right)
\end{equation}
where $\bar{r}_i = \sum_{h=1}^t r_{ih}/t$ is the sample mean. The related Pearson correlation coefficient is defined as
\begin{equation}
c_{ij} = \frac{\sigma_{ij}}{\sqrt{\sigma_{ii}\,\sigma_{jj}}}
\end{equation}
\subsubsection{Hierarchical Clustering Average Linkage (HCAL)}

The hierarchical clustering is an agglomerative algorithm that recursively clusters groups of objects according to a distance. The latter is defined in the simplest way in Ref.\cite{bongiorno2019nested}: the Pearson correlation matrix $C$ is transformed into a distance matrix $D$ as follows
\begin{equation}
d_{ij} = 1 - c_{ij},
\end{equation} 
which respects the axioms of a distance.
Then a distance metric among clusters must by defined: in the HCAL case, it is based on the average linkage between clusters $p$ and $q$
\begin{equation}
\rho_{pq} = \frac{\sum_{i \in \mathfrak{C}_p} \sum_{j \in \mathfrak{C}_q} d_{ij}}{n_q\,n_p},\label{eq:rho_pq}
\end{equation}
where $\mathfrak{C}_p$ and $\mathfrak{C}_q$ are the sets of elements belonging to the clusters $p$ and $q$ respectively, and $n_p$ and $n_q$ are their cardinality.

Hierarchical clustering works as follows: initially, each element has its own cluster. Then, the pair of clusters $(p,q)$ with the smallest distance $\rho_{pq}$ are merged together into a new cluster $s$ such that $\mathfrak{C}_s = \mathfrak{C}_p \cup \mathfrak{C}_q$. The algorithm recursively joins a pair of clusters until all nodes fall into a single unique cluster. The genealogy $\mathfrak{G}$ of the hierarchical clustering can be uniquely identified by the sequence of $n-2$ joins among the pairs of clusters identified by the method, and this defines a dendrogram.

\subsubsection{The Filtered Matrix}\label{sec:ALCA}
Ref.~\cite{tumminello2007hierarchically} proposes to clean the correlation sub-matrix defined from the indices $\mathfrak{F}_{pq} = \{ (i,j)\,:\, i \in \mathfrak{C}_p,\, j \in \mathfrak{C}_q\}$ by replacing all its elements with their average: mathematically one builds a matrix $C^<$ with elements
\begin{equation}
c_{ij}^< = c_{ji}^<=1 - \rho_{pq}  \;\; \mbox{where} \;\;\;  (p,q ) \in  \mathfrak{G},\,  (i,j) \in \mathfrak{F}_{pq},
\end{equation}
$\rho_{pq}$ is the average distance between clusters $p$ and $q$ (see \eqref{eq:rho_pq}) and the diagonal of $C^<$ is set to 1. An equivalent description of this approach is in terms of the factor loading matrix, as in the original paper~\cite{tumminello2007hierarchically}. 

It is important to stress that the matrix $C^<$ will be positively defined by construction \cite{tumminello2007hierarchically}. The main feature of this model is to obtain the simplest matrix $C^<$ that shares the same dendrogram as $C$; this means that by applying the HCAL to both $C$ and $C^<$, the resulting dendrograms will be identical. However, we believe that this is also one of the main limitations of this approach; in fact, it does not account for the presence of overlap among clusters.

\subsection{Bootstrap Average Linkage Correlation Matrix}\label{sec:BAHC}
To overcome these two issues of HCAL filtering while keeping its advantages, we propose here a new approach to filter correlation matrices based on data matrix bootstrap resampling of the feature indices; therefore, it better accounts for the influence of randomness on the inferred structure. We call it BAHC, which stands for Bootstrap-averaged hierarchical clustering.

Our recipe prescribes to create a set of $m$ bootstrap copies of the data matrix $R$, denoted by $\{R^{(1)},\, R^{(2)},\, \cdots, R^{(m)}\}$. A single bootstrap copy of the data matrix $R^{(b)} \in \mathbb{R}^{n \times t}$ is defined entry-wise as $r^{(b)}_{ij} = r_{i s^{(b)}_j}$, where ${\bf{s}}^{(b)}$ is a vector of dimension $t$ obtained with random sampling by replacement of the elements of the vector $\{1, 2, \cdots, t\}$. The vector ${\bf{s}}^{(b)}$, $b=1,\cdots,m$ are independently sampled.

Each bootstrap copy $b$ of the data matrix has an associated Pearson correlation matrix $C^{(b)}$ from which we can construct the HCAL filtered matrix $C^{(m)<}$. Finally, each element of the filtered Pearson correlation matrix $C^{\textrm{BAHC}}$ is defined as the average over the $m$ filtered bootstrap copies, i.e.,
\begin{equation}\label{eq:corfilt}
c_{ij}^{\textrm{BAHC}} = \sum_{b=1}^m \frac{c_{ij}^{(b)<}}{m}
\end{equation}
We  stress that since $C^{(h)<}$ are  positive define matrices by construction, $C^{\textrm{BAHC}}$ is also a   positive defined matrix.

The main advantage of the BAHC method is not to force $C^{\textrm{BAHC}}$ to be embedded in a purely hierarchical structure. Indeed, different bootstraps may yield different dendrograms, in which case a strict hierarchical structure is too stringent. Thus, the BAHC method can reproduce some degree of overlap among clusters defined in a hierarchical way.

\subsection{Filter Covariance Matrices}\label{sec:COV}
To build BAHC-filtered covariance matrices, we first estimate bootstrapped univariate variances
$\{\sigma_{ii}^{(1)},\, \sigma_{ii}^{(2)},\, \cdots, \sigma_{ii}^{(m)}\}$, where a generic element $\sigma_{ii}^{(b)}$ of $\Sigma^{(b)}$ is defined as 
\begin{equation}
\sigma_{ii}^{(b)} = \frac{1}{t} \sum_{h=1}^t \left( r^{(b)}_{ij} - \bar{r}_i^{(b)} \right)^2 
\end{equation}

Then element  $(i,j)$ of  $b$-th bootstrap covariance is defined as
\begin{equation}
\sigma_{ij}^{(b)<} = c_{ij}^{(b)<}\, \sqrt{\sigma_{ii}^{(b)}\, \sigma_{jj}^{(b)} } 
\end{equation}

Finally, as in \eqref{eq:corfilt}, the element $(i,j)$ of the filtered covariance matrix is defined as
\begin{equation}\label{eq:covfilt}
\sigma_{ij}^{\textrm{BAHC}} = \sum_{h=1}^m \frac{\sigma_{ij}^{(h)<}}{m}
\end{equation}

\section{Eigenvector in- and out-of-sample overlap from the Oracle estimator}

We recall the concept of Oracle estimator $\Xi$: given the spectral decomposition of the in-sample correlation matrix $C^{in} = U^{in} \Lambda^{in} U^{in\dagger}$ and the spectral decomposition of the out-of-sample correlation matrix $C^{out} = U^{out} \Lambda^{out} U^{out\dagger}$, where $\Lambda^{in/out}$ are diagonal eigenvalue matrices made from the eigenvalues of $C^{in/out}$, and $U^{in/out}$ is the matrix defined by the eigenvectors of $C^{in/out}$, the Oracle eigenvalue matrix is defined as 
\begin{equation}
Z^{in} = \left( U^{in\dagger} C^{out} U^{in} \right)_d
\end{equation}
where the superscript  $in$ indicates that we used the in-sample eigenvectors for its estimation. The operator $\left(  \right)_d$ sets to zero all the off-diagonal elements. Then the Oracle estimator of the correlation matrix is defined as
\begin{equation}
\Xi^{in} = U^{in} Z^{in} U^{in \dagger}.
\end{equation}
Ref.~\cite{bun2016rotational} shows that Oracle eigenvalues are the optimal correction of the in-sample eigenvalues $\Lambda^{in}$ in the sense that it minimizes the Frobenius norm of the difference between the out-of-sample correlation matrix and the corrected in-sample one $\left\lVert C^{out} - \Xi^{in} \right\rVert _F$. Although this estimator sounds worryingly tautological, since it require the knowledge of the out-of-sample correlation to construct the most similar estimator, Ref.~\cite{bun2016rotational} show that is possible to obtain $Z^{in}$ in the $t,n\to\infty$ at constant $q=n/t$ limit without the knowledge of $C^{out}$ for a broad set of distributions and noises (multiplicative and additive) if the system is stationary and for $t>n$ (low-dimensional regime). Indeed, it easy to show that the Oracle estimator is exactly $C^{out}$ if and only if $U^{in} = U^{out}$ since 
\begin{eqnarray}
Z^{in} = \left( U^{in\dagger} C^{out} U^{in} \right)_d = \left( U^{in\dagger}U^{out} \Lambda^{out} U^{out\dagger} U^{in} \right)_d =  \nonumber \\
= \left( U^{out\dagger}U^{out}  \Lambda^{out} U^{out\dagger}U^{out} \right)_d =  \left(  \Lambda^{out} \right)_d = \Lambda^{out}.
\end{eqnarray}

Therefore the Frobenius norm of $\left\lVert \Xi^{in} - C^{out} \right\rVert _F$ can be interpreted as a measure of the overlap between the out-of-sample eigenvectors $U^{out}$ and the in-sample ones $U^{in}$.

\section{Global minimum-variance portfolios in other equity markets}
\begin{figure}
\includegraphics[width=0.5\columnwidth]{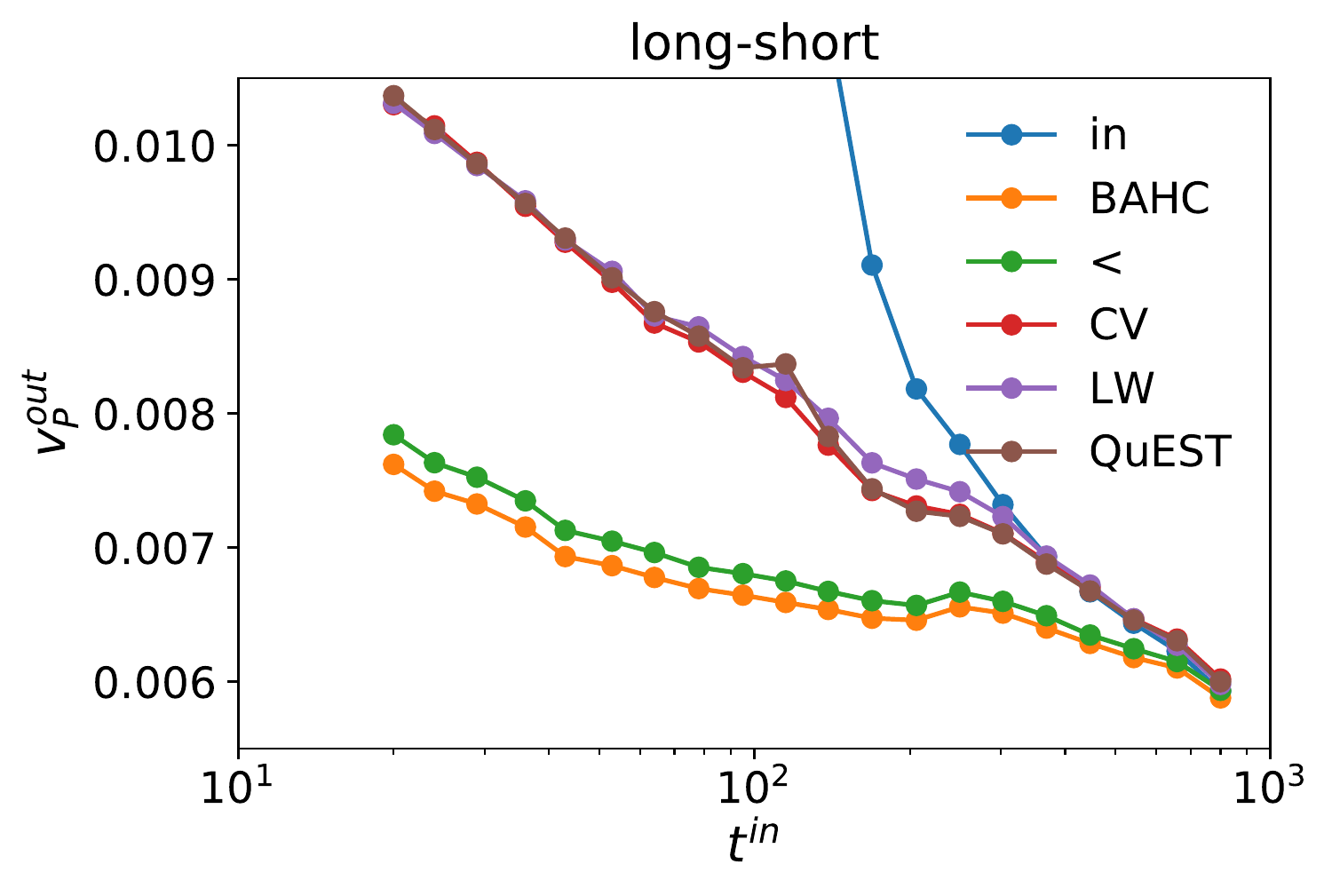}\includegraphics[width=0.5\columnwidth]{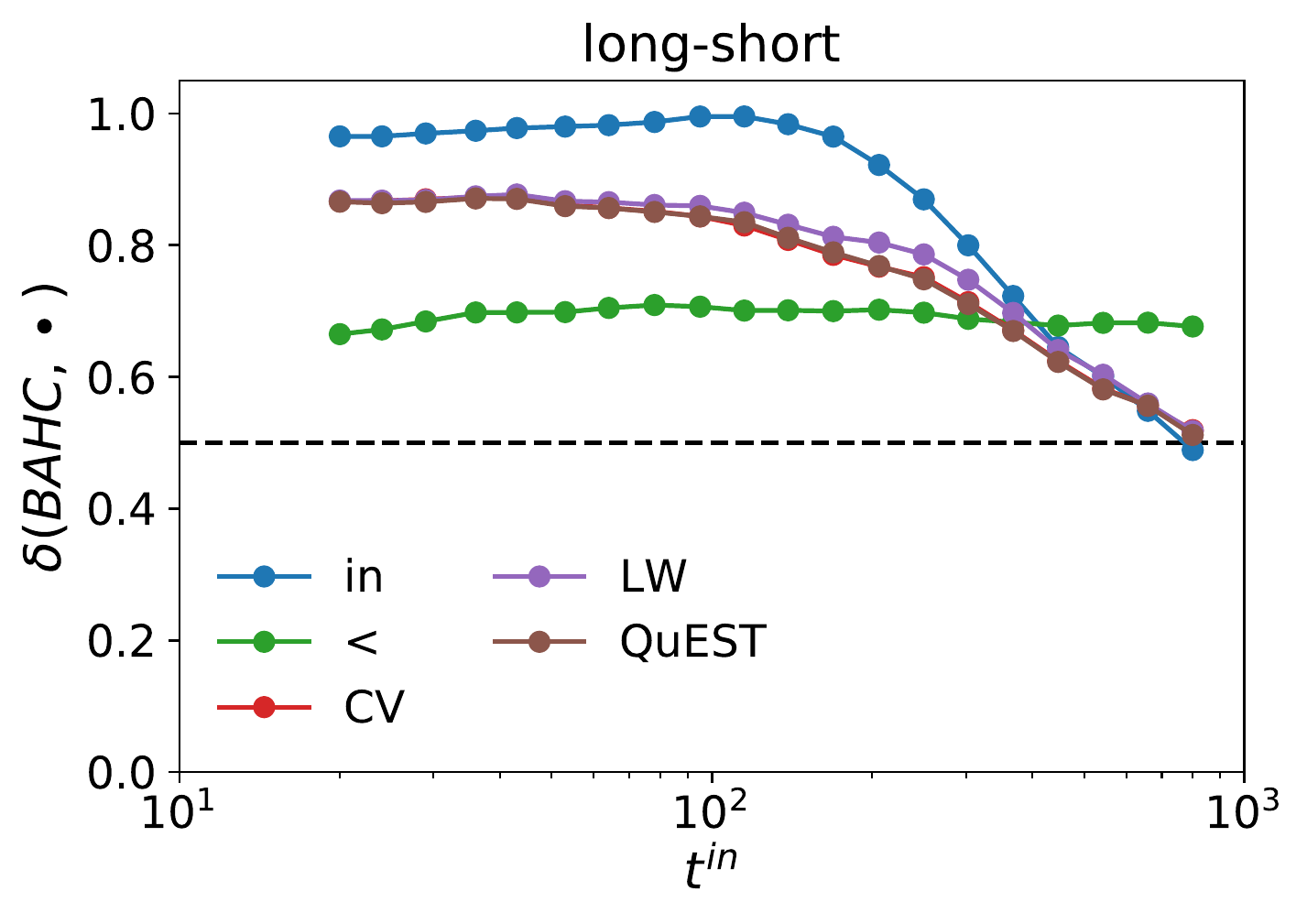}

\includegraphics[width=0.5\columnwidth]{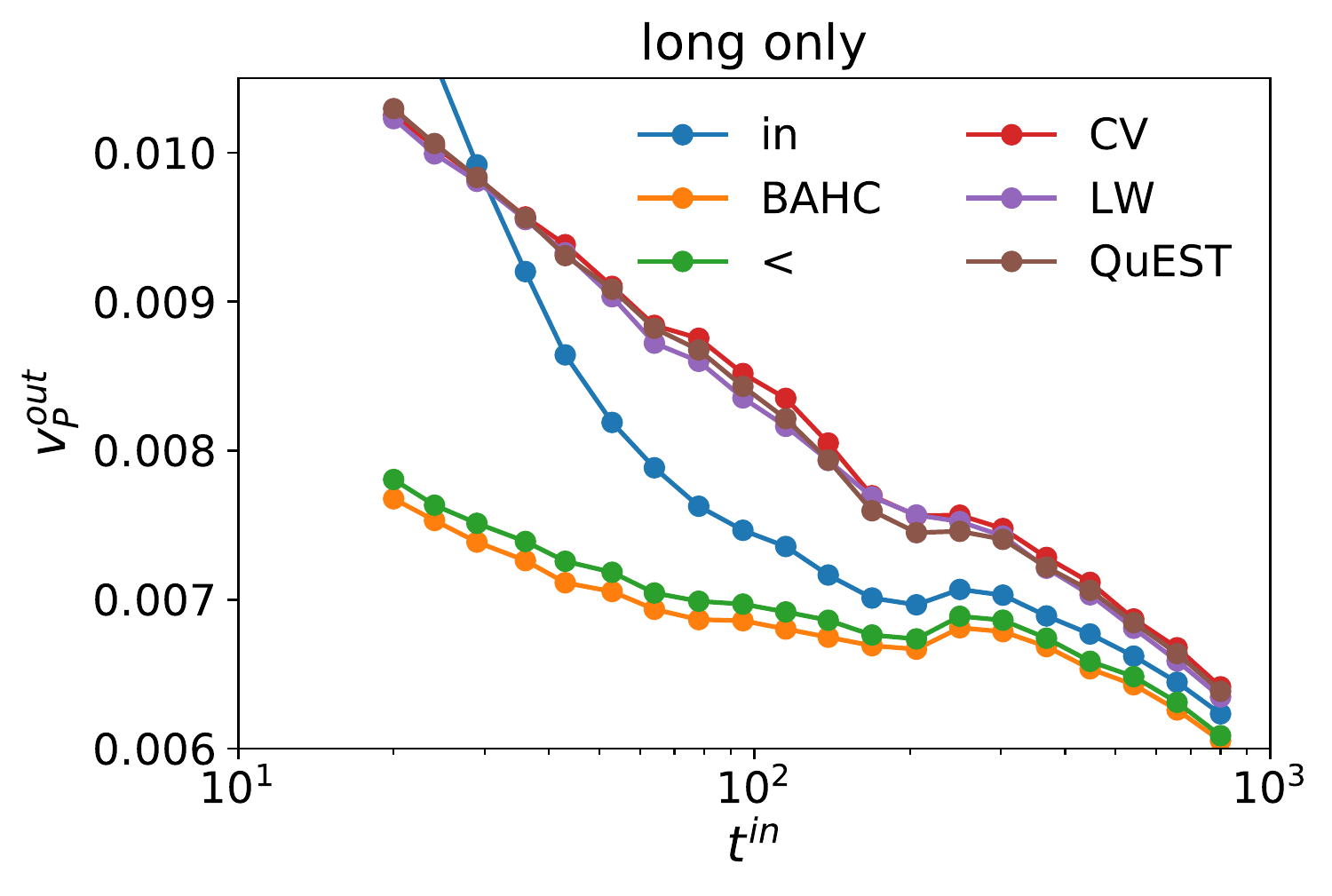}
\includegraphics[width=0.5\columnwidth]{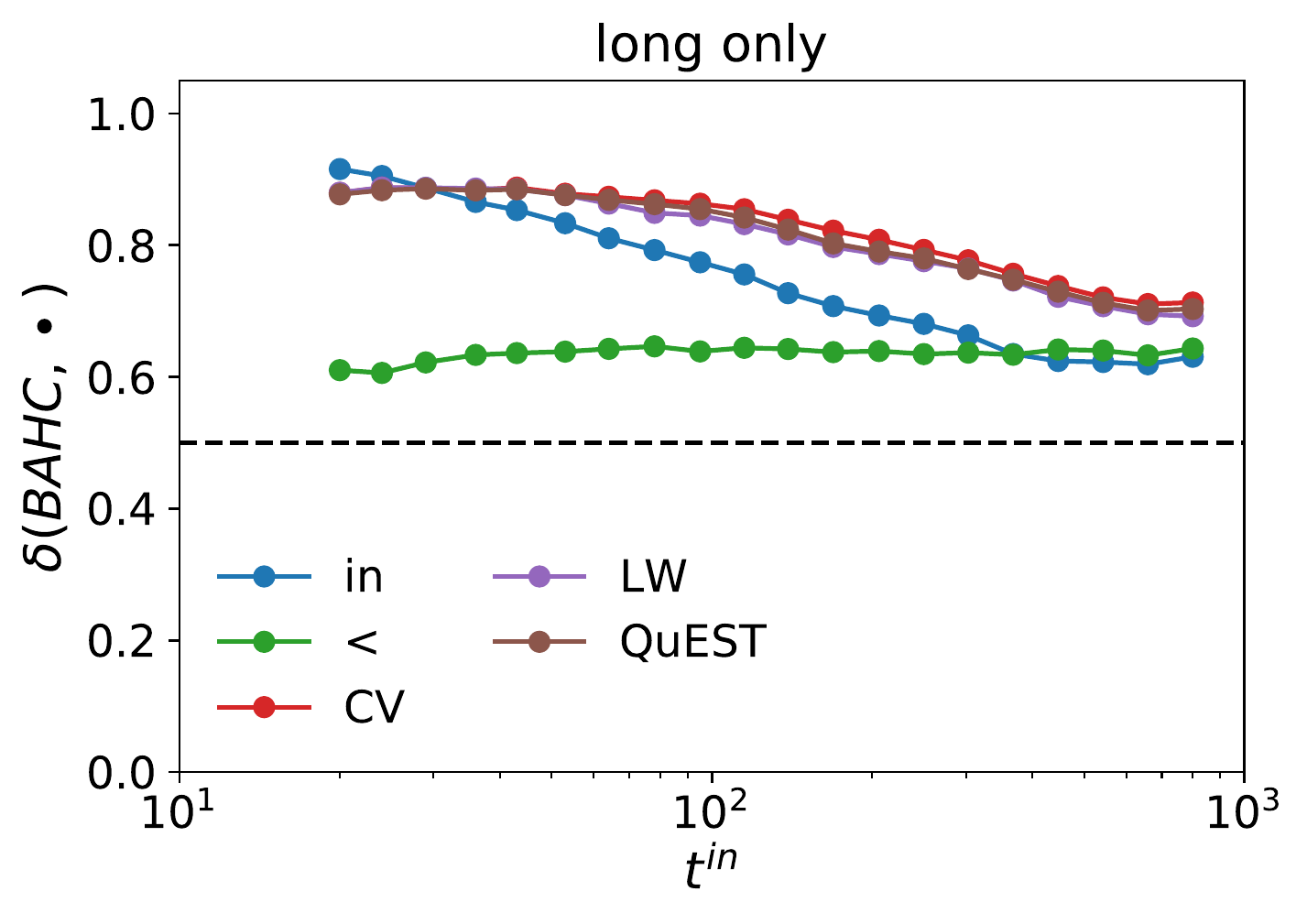}

\caption{Left plots: realized risk for different estimators; right plots: fraction of time the realized risk of BAHC is smaller than the one obtained with alternative estimators.  $10,000$ independent simulations per point; $t^{out}=42$ days, $n=100$ assets, Hong Kong equities. \label{fig:hongkong}}
\end{figure}

\begin{figure}

\includegraphics[width=0.5\columnwidth]{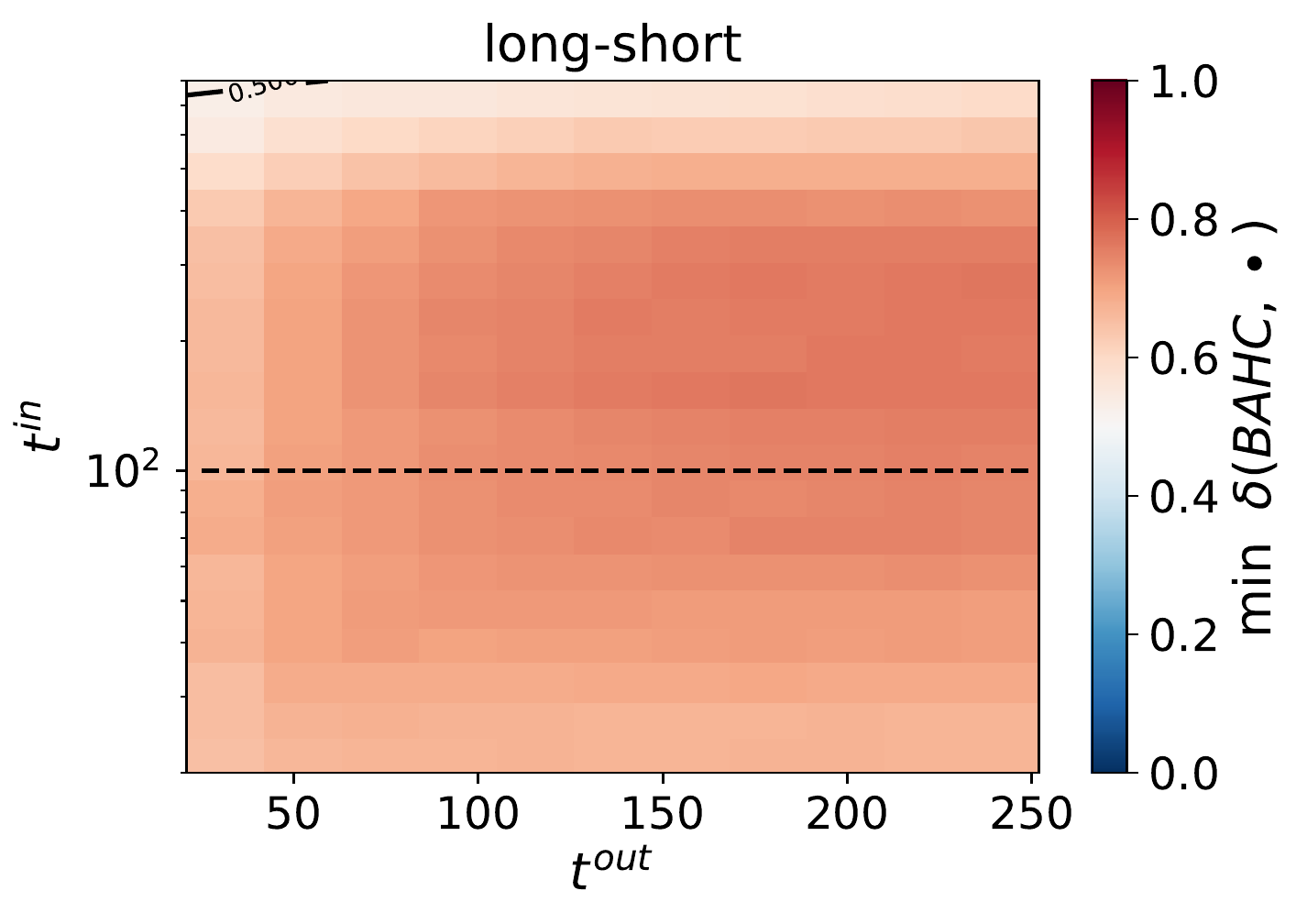}\includegraphics[width=0.5\columnwidth]{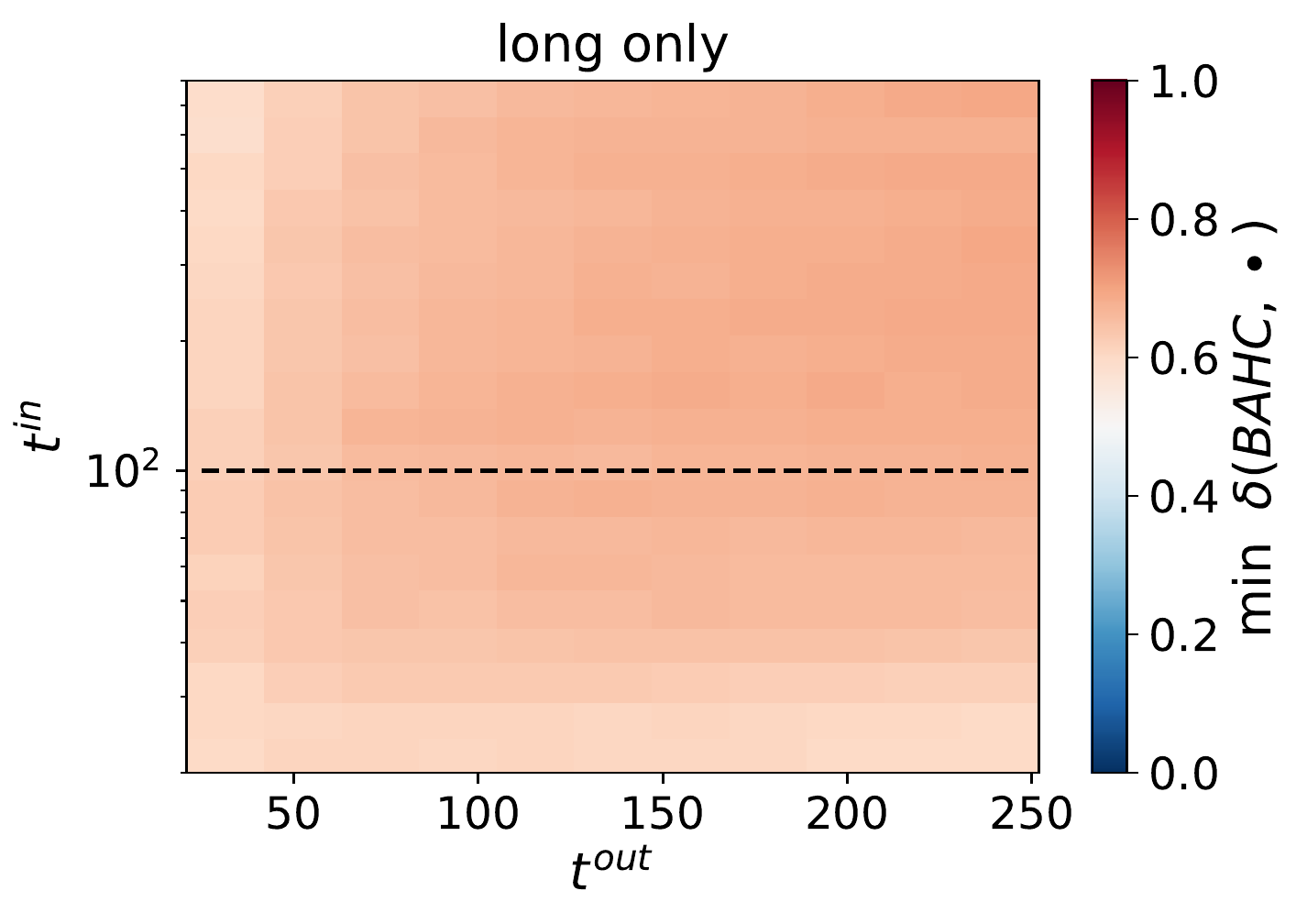}

\caption{Fraction of time the realized risk of BAHC is smaller than the best performing alternative method.  Left plot: portfolios with long and short positions, right plot: portfolios with only long positions. The level curve corresponds to a 50\% probability. $10,000$ independent simulations per point; $n=100$ assets\label{fig:hongkong_tout}}
\end{figure}

Figures \ref{fig:hongkong} and \ref{fig:hongkong_tout} report the out-of-sample risk of covariance matrix cleaning methods with the same set up for Hong Kong stock exchange \ref{fig:hongkong}. The analysis cover $1281$ stocks in from 2005-10-19 to 2017-06-23. The stocks are not simultaneously listed over all time-period: the number stocks ranges from $590$ on 2008-08-22 to $1277$ on 2017-06-14. Results are qualitatively consistent with those observed in the US equity market.

\section{List of large-capitalization assets in the US equities dataset}

{\tiny A, AA, AAN, AAP, AAPL, ABC, ABT, ACGL, ACM, ACN, ACV, ADBE, ADI, ADM, ADP, ADS, ADSK, AEE, AEO, AEP, AES, AET, AFG, AFL, AGCO, AGN, AGO, AHL, AIG, AIV, AIZ, AJG, AKAM, AKS, ALB, ALEX, ALL, ALTR, ALV, ALXN, AMAT, AMD, AME, AMG, AMGN, AMP, AMR, AMT, AMTD, AMZN, AN, ANAT, ANF, ANSS, AON, APA, APC, APD, APH, ARCC, ARE, ARW, ASH, ATI, ATLS, ATO, ATR, ATVI, AVB, AVGO, AVP, AVT, AVX, AVY, AWI, AWK, AXP, AXS, AZO, BA, BAC, BAX, BBBY, BBT, BBY, BDN, BDX, BEN, BG, BIG, BIIB, BIO, BJ, BK, BKD, BLK, BLL, BMRN, BMS, BMY, BOH, BOKF, BPOP, BR, BRO, BSX, BTU, BWA, BXP, BXS, C, CA, CAG, CAH, CAL, CAT, CB, CBS, CBSH, CBT, CCE, CCI, CCK, CCL, CCO, CDNS, CE, CECO, CELG, CERN, CETV, CF, CFFN, CFR, CHD, CHH, CHK, CHRW, CHS, CI, CIEN, CIM, CINF, CIT, CKH, CL, CLB, CLF, CLGX, CLI, CLR, CLX, CMA, CMC, CMCSA, CME, CMG, CMI, CMP, CMS, CNA, CNP, CNX, COF, COG, COL, COO, COP, COST, CPA, CPB, CPRT, CPT, CPWR, CR, CREE, CRK, CRL, CRM, CRS, CSCO, CSL, CSX, CTAS, CTL, CTSH, CTV, CTXS, CVA, CVG, CVS, CVX, CXO, CXW, CY, CYH, D, DAL, DBD, DCI, DDR, DE, DEI, DF, DFS, DG, DGX, DHI, DHR, DIS, DISCA, DISH, DKS, DLB, DLR, DLTR, DNB, DNR, DO, DOV, DOX, DPS, DRE, DRI, DTE, DTV, DUK, DVA, DVN, EAT, EBAY, ECL, ED, EFX, EGN, EIX, EL, EMN, EMR, ENDP, ENR, EOG, EQIX, EQR, EQT, ESRX, ESS, ETN, ETR, EV, EW, EWBC, EXC, EXP, EXPD, EXPE, F, FAST, FCN, FCX, FDS, FDX, FE, FFIV, FHN, FII, FIS, FISV, FITB, FL, FLIR, FLO, FLR, FLS, FMC, FNF, FOSL, FRO, FRT, FSLR, FTI, FTR, FULT, G, GCI, GD, GDI, GE, GEF, GES, GGG, GGP, GHL, GILD, GIS, GLW, GME, GNTX, GNW, GOOG, GPC, GPN, GPRO, GPS, GRMN, GS, GT, GWW, H, HAL, HAS, HBAN, HBI, HCC, HCP, HD, HE, HES, HI, HIG, HK, HLF, HOG, HOLX, HON, HP, HPQ, HPT, HRB, HRC, HRL, HRS, HSC, HSIC, HST, HSY, HTZ, HUM, HUN, IBKR, IBM, ICE, IDXX, IEX, IFF, IGT, ILMN, INTC, INTU, IP, IPG, IPI, IR, IRM, ISCA, ISRG, IT, ITRI, ITT, ITW, IVZ, JBHT, JBL, JCI, JCP, JEC, JEF, JLL, JNJ, JNPR, JOE, JPM, JWN, K, KAR, KBH, KBR, KEX, KEY, KIM, KLAC, KMB, KMT, KMX, KO, KR, KSS, KSU, L, LAMR, LAZ, LBTYA, LEA, LECO, LEG, LEN, LH, LIFE, LII, LLL, LLY, LM, LMT, LNC, LNT, LOW, LPNT, LRCX, LSI, LSTR, LUV, LVS, M, MA, MAC, MAN, MAR, MAS, MAT, MBI, MCD, MCHP, MCK, MCO, MCY, MD, MDC, MDP, MDR, MDRX, MDT, MDU, MET, MGM, MHK, MKC, MKL, MLM, MMC, MMM, MO, MORN, MOS, MRK, MRO, MRVL, MS, MSFT, MSM, MTB, MTD, MTW, MU, MUR, MXIM, MYGN, MYL, NATI, NAV, NBL, NBR, NCR, NDAQ, NEE, NEM, NFG, NFLX, NFX, NI, NIHD, NKE, NLY, NOC, NOV, NRG, NSC, NSM, NTAP, NTRS, NUAN, NUE, NVDA, NVR, NWL, NWSA, NYT, O, OC, ODP, OFC, OGE, OI, OII, OIS, OKE, OMC, ORA, ORCL, ORI, ORLY, OSK, OXY, PAYX, PBCT, PBI, PCAR, PCG, PDCO, PDM, PEG, PENN, PEP, PFE, PFG, PG, PGR, PH, PHM, PKG, PKI, PLD, PM, PNC, PNR, PNW, PPG, PPL, PRGO, PRU, PSA, PTEN, PVH, PWR, PX, PXD, QCOM, R, RBC, RCL, RDC, RE, REG, REGN, RF, RGA, RGLD, RHI, RHT, RJF, RL, RMBS, RMD, RNR, ROK, ROP, ROST, RPM, RRC, RRD, RS, RSG, RTN, RYN, S, SATS, SBAC, SBUX, SCCO, SCG, SCHN, SCHW, SCI, SD, SE, SEE, SEIC, SHLD, SHW, SIG, SIRI, SJM, SLAB, SLB, SLG, SLM, SM, SMG, SNA, SNH, SNPS, SNV, SO, SON, SPG, SPN, SPR, SRCL, SRE, STI, STLD, STRA, STT, STX, STZ, SUN, SVU, SWK, SWKS, SWN, SYK, SYMC, SYY, T, TAP, TCO, TDC, TDG, TDS, TDW, TECD, TECH, TER, TEX, TFSL, TFX, TGT, THC, THG, THO, TIF, TJX, TK, TKR, TMK, TMO, TOL, TPX, TRI, TRMB, TRN, TROW, TRV, TSCO, TSN, TSS, TTC, TUP, TXN, TXT, UDR, UFS, UGI, UHS, UNH, UNM, UNP, UNT, UPL, UPS, URBN, USB, USG, USM, UTHR, UTX, V, VAR, VFC, VLO, VLY, VMC, VMI, VMW, VNO, VR, VRSK, VRSN, VRTX, VRX, VSH, VTR, VVC, VZ, WAB, WAT, WBC, WCC, WCN, WDC, WDR, WEC, WEN, WFC, WFT, WHR, WIN, WLL, WM, WMB, WMS, WMT, WRB, WRI, WSC, WSM, WTM, WTR, WTW, WU, WY, WYNN, X, XEC, XEL, XL, XLNX, XOM, XRAY, XRX, Y, YUM, ZBRA, ZION
}

\end{document}